\documentclass[a4paper,11pt]{article}
\pdfoutput=1

\usepackage{multirow}
\usepackage{scalerel}
\usepackage{jcappub}
\usepackage[T1]{fontenc}
\usepackage[utf8]{inputenc}
\usepackage{mathrsfs}
\usepackage{amsmath,amssymb}
\usepackage[normalem]{ulem}
\usepackage{amssymb}
\usepackage{diagbox}
\usepackage{float}
\usepackage{slashed}
\usepackage{{bbold}}
\newcommand{\fortepiano}{\texttt{FortEPiaNO}}
\newcommand{\Neff}{\ensuremath{N_{\rm eff}}}
\usepackage{siunitx}
\usepackage{graphicx}
\usepackage{subcaption}

\usepackage{todonotes}

\usepackage{tikz}

\usetikzlibrary{external}

\usetikzlibrary{patterns}
\usetikzlibrary{calc,arrows,positioning}
\usetikzlibrary{arrows,shadows}
\usetikzlibrary{decorations.pathmorphing}	
\usetikzlibrary{decorations.markings}

\tikzset{
>=latex, 
vector/.style={decorate, decoration={snake,amplitude=2pt,segment length=1.7mm}, draw},
fermion/.style={draw=black, postaction={decorate},
decoration={markings,mark=at position .55 with {\arrow[draw=black]{>}}}},
fermionbar/.style={draw=black, postaction={decorate},
decoration={markings,mark=at position .55 with {\arrow[draw=black]{<}}}},
fermionnoarrow/.style={draw=black},
}

\title{\boldmath Towards a precision calculation of $\Neff$ in the Standard Model II: Neutrino decoupling in the presence of flavour oscillations and finite-temperature QED
}
\author[a]{Jack~J.~Bennett,}
\author[b]{Gilles~Buldgen,}
\author[c]{Pablo F.~de~Salas,}
\author[b]{Marco~Drewes,}
\author[d,e]{Stefano~Gariazzo,}
\author[e]{Sergio~Pastor,}
\author[a]{and Yvonne~Y.~Y.~Wong}

\affiliation[a]{Sydney Consortium for Particle Physics and Cosmology, School of Physics, The University of New South Wales, Sydney NSW 2052, Australia}
\affiliation[b]{Centre for Cosmology, Particle Physics and Phenomenology,
Universit\'{e} catholique de Louvain, Chemin du cyclotron, 2,
Louvain-la-Neuve B-1348, Belgium}
\affiliation[c]{The Oskar Klein Centre for Cosmoparticle Physics,
Department of Physics, Stockholm University, SE-106 91 Stockholm, Sweden}
\affiliation[d]{INFN, Sezione di Torino, Via P. Giuria 1, I--10125 Torino, Italy}
\affiliation[e]{Instituto de F{\'\i}sica Corpuscular (CSIC-Universitat de Val{\`e}ncia),
Parc Cient{\'\i}fic UV, C/ Ca\-te\-dr{\'a}tico Jos{\'e} Beltr{\'a}n, 2, E-46980 Paterna (Valencia), Spain}

\emailAdd{j.j.bennett@unsw.edu.au, gilles.buldgen@uclouvain.be, pablo.fernandez@fysik.su.se, marco.drewes@uclouvain.be, gariazzo@to.infn.it, pastor@ific.uv.es, yvonne.y.wong@unsw.edu.au}

\date{\today}

\abstract{We present in this work a new calculation of the standard-model benchmark value for the effective number of neutrinos, $\Neff^{\rm SM}$, that quantifies the cosmological neutrino-to-photon energy densities.
The calculation takes into account neutrino flavour oscillations, finite-temperature effects in the quantum electrodynamics plasma to ${\cal O}(e^3)$, where $e$ is the elementary electric charge, and a full evaluation of the neutrino--neutrino collision integral.
We provide furthermore a detailed assessment of the uncertainties in the benchmark $\Neff^{\rm SM}$ value, through testing the value's dependence on (i)~optional approximate modelling of the weak collision integrals, (ii)~measurement errors in the physical parameters of the weak sector, and (iii)~numerical convergence, particularly in relation to momentum discretisation.
Our new, recommended standard-model benchmark is $\Neff^{\rm SM} = 3.0440 \pm 0.0002
$, where the nominal uncertainty is attributed predominantly to errors incurred in the numerical solution procedure ($|\delta \Neff| \sim10^{-4}$), augmented by measurement errors in the solar mixing angle $\sin^2\theta_{12}$ ($|\delta \Neff| \sim10^{-4}$).}

\begin{document}

\begin{flushright}
{\large \tt CPPC-2020-10}
\end{flushright}	

\maketitle

\section{Introduction}\label{sec:intro}

The effective number of neutrinos, $\Neff$, is a parameter that quantifies the ratio of cosmological energy densities in neutrino-like relics%
\footnote{In the cosmological context, a neutrino-like relic is a thermalised or partially thermalised light particle state that decouples from the primordial plasma while ultra-relativistic and remains decoupled thereafter.}
to photons in the early universe shortly after neutrino decoupling (at temperatures $T \sim 1$~MeV).
Within the standard model (SM) of particle physics,%
\footnote{In this work, the term ``standard model of particle physics'' is understood to include neutrino oscillations.}
its expected theoretical value, $\Neff^{\rm SM}$, is $3$ (for three generations of SM neutrinos), plus percent-level corrections due primarily to (i) energy transport from the primordial quantum electrodynamics (QED) plasma to the neutrino sector~\cite{Dodelson:1992km,Hannestad:1995rs,Dolgov:1997mb,Dolgov:1998sf,Esposito:2000hi,Mangano:2005cc,Birrell:2014uka,Grohs2016,Escudero:2018mvt,Escudero:2020dfa,Froustey:2019owm,Akita:2020szl,Froustey:2020mcq}, and (ii) finite-temperature-induced deviations of the QED plasma equation of state from an ideal gas~\cite{Dicus:1982bz,Heckler:1994tv,Fornengo:1997wa,Lopez:1998vk,Mangano:2001iu,Bennett:2019ewm}.
A 2016, fully momentum-dependent precision transport study including neutrino flavour oscillations and finite-temperature QED corrections performed by some of us (de Salas and Pastor) put the number at $\Neff^{\rm SM} = 3.045$~\cite{deSalas:2016ztq}.
This number has recently been revised to $\Neff^{\rm SM} = 3.044$ in references~\cite{Akita:2020szl,Froustey:2020mcq},  including a new QED correction identified in reference~\cite{Bennett:2019ewm}.

Interest in the $\Neff$ parameter as a probe of beyond-the-standard model (BSM) light relics --- and by association the precision computation of $\Neff^{\rm SM}$ --- has a long history. In the era of precision cosmology, the continued interest in the $\Neff$ parameter stems primarily from its effects on observables such as the cosmic microwave background (CMB) anisotropies; current measurements by the ESA Planck mission already constrain the parameter to $\Neff=2.99^{+0.34}_{-0.33}$~(95\% C.I.)~\cite{Aghanim:2018eyx},
severely limiting the viable parameter space of BSM states produced well after the quantum chromodynamics (QCD) phase transition ($T \sim 200$~MeV) such as light sterile neutrinos~\cite{Hannestad:2015tea,Hagstotz:2020ukm}, eV-mass axions~\cite{Archidiacono:2013cha}, and decay products of long-lived BSM particles~\cite{Hasenkamp:2012ii,DiBari:2013dna}.
In the near future, a conservative projection of the scientific capacity of CMB-S4 sees the 1$\sigma$ sensitivity to $\Neff$ improve to $\sigma(N_{{\rm eff}})\sim 0.02 \to 0.03$~\cite{Abazajian:2016yjj} depending on the final configuration of the experiment, potentially probing light relic production immediately after the QCD phase transition~\cite{Archidiacono:2015mda,Abazajian:2019oqj}.

Increasing observational precision calls for a corresponding improvement in the theoretical prediction of the SM benchmark $\Neff^{\rm SM}$. Recently, three of us (Gariazzo, de Salas, and Pastor) revisited the energy transport aspect of the $\Neff^{\rm SM}$ computation in the presence of neutrino flavour oscillations~\cite{Gariazzo:2019gyi}, while the
other four (Bennett, Buldgen, Drewes and Wong) quantified the leading and several sub-leading finite-temperature QED contributions to $\Neff^{\rm SM}$~\cite{Bennett:2019ewm}.
In this work, we pool our resources to supply a state-of-the-art precision computation of the SM benchmark $\Neff^{\rm SM}$. Relative to the 2016 update of~\cite{deSalas:2016ztq}, the main improvements in this new calculation are:
\begin{enumerate}
\setlength{\itemsep}{-0.0em}
\item Energy transport in the presence of flavour oscillations is now computed using the new precision neutrino decoupling code~\fortepiano~\cite{Gariazzo:2019gyi}\footnote{This code will be made publicly available at \url{https://bitbucket.org/ahep_cosmo/fortepiano_public}.},
a fully momentum-dependent decoupling code that accepts up to 3 active + 3 sterile neutrino flavours.

\item We incorporate two new, subdominant finite-temperature corrections to the QED equation of state identified in~\cite{Bennett:2019ewm}, notably the ${\cal O}(e^3)$ correction, where $e$ is the elementary electric charge. As we shall see, this correction will account for the leading-order shift in the SM benchmark $\Neff^{\rm SM}$ relative to previously reported results.

\item We include a complete numerical evaluation of the neutrino--neutrino collision integral in the presence of neutrino flavour oscillations. Of particular note are the flavour-off-diagonal entries peculiar to oscillating systems: these had previously been modelled using a damping approximation in references~\cite{deSalas:2016ztq,Gariazzo:2019gyi}.

\item We provide a detailed assessment of the uncertainties in the benchmark $\Neff^{\rm SM}$ value, through testing the value's dependence on (i)~the approximate modelling of the weak collision integrals in the presence of flavour oscillations, (ii)~measurement errors in the physical parameters of the weak sector, and (iii)~numerical convergence.

\end{enumerate}

For the reader in search of a quick number, our new SM benchmark is
$\Neff^{\rm SM}=3.0440\pm 0.0002
$, where the nominal uncertainty is attributed predominantly to errors incurred in the numerical solution procedure ($|\delta \Neff|\sim10^{-4}$), followed by measurement uncertainties in the solar mixing angle $\sin^2\theta_{12}$ ($|\delta \Neff|\sim10^{-4}$). This new benchmark value is identical to that obtained in reference~\cite{Froustey:2020mcq} to the same number of significant digits, which had also been computed with the full neutrino--neutrino collision integral and finite-temperature QED effects to ${\cal O}(e^3)$.
For readers interested in the details of our calculation, the rest of the paper is organised as follows.
We describe the physical system, including the relevant equations of motion, in section~\ref{sec:system}, and the new elements of the present calculation in section~\ref{sec:whatsnew}. Section~\ref{sec:convergence} discusses the issue of numerical convergence with respect to the initialisation and momentum discretisation procedures, in terms of controlled tests of detailed balance.
Our calculations of $\Neff^{\rm SM}$ under the introduction of various new elements are presented in section~\ref{sec:results}.
Section~\ref{sec:conclusions} contains our conclusions. Two appendices document new technical details of our calculations.


\section{The system}\label{sec:system}

The SM benchmark effective number of neutrinos $\Neff^{\rm SM}$ is defined via the ratio of the neutrino energy density $\rho_{\nu}$ to the photon energy density $\rho_\gamma$ in the limit $T/m_e \to0$,
\begin{equation}
\label{Neff_def}
\left. \frac{\rho_\nu}{\rho_\gamma} \right|_{T/m_{e} \to 0} \equiv \frac{7}{8} \Big(\frac{4}{11}\Big)^{4/3} \Neff^{\rm SM},
\end{equation}
where $T$ is the photon temperature and $m_e$ is the electron mass. Its precision computation typically requires that we solve two sets of equations of motion across the relic neutrino decoupling epoch,
in a time frame corresponding to $T \sim {\cal O}(100) \to {\cal O}(0.01)$~MeV.
These two sets of equations are:
(i) the continuity equation, which tracks the total energy density of the universe in the time frame of interest, and (ii) the generalised Boltzmann equation --- also known as the quantum kinetic equation --- for the density matrix of the neutrino ensemble, which follows the non-equilibrium dynamics of the ensemble in the presence of flavour oscillations and particle scattering.


\subsection{Continuity equation}

In a Friedmann--Lema\^{\i}tre--Robertson--Walker (FLRW) universe, the continuity equation reads $({\rm d}/{\rm d}t) \rho_{\rm tot} + 3 H (\rho_{\rm tot} + P_{\rm tot}) = 0$,
where $\rho_{\rm tot}$ and $P_{\rm tot}$ are, respectively, the total {\it physical} energy density and pressure, $H\equiv (1/a) ({\rm d} a/{\rm d} t)$ the Hubble expansion rate, and $a$ is the scale factor. In terms of the rescaled time, momentum, and temperature coordinates, $x \equiv a m_e$, $y \equiv a p$, and $z \equiv aT_\gamma $, the continuity equation is given equivalently by~(e.g.,~\cite{Mangano:2001iu})
\begin{align}
\frac{{\rm d}}{{\rm d} x} \bar{\rho}_{\rm tot}(x,z(x))=\frac{1}{x}\left[\bar{\rho}_{\rm tot}(x,z(x))-3\bar{P}_{\rm tot}(x,z(x))\right], \label{comoving_conserv_equation}
\end{align}
where $\bar{\rho}_{\rm tot} \equiv (x/m_e)^4 \rho_{\rm tot}$ and $\bar{P}_{\rm tot} \equiv (x/m_e)^4 P_{\rm tot}$ are now the {\it comoving} quantities.
Thus, in using rescaled and comoving variables, we have factored out the effects of cosmic expansion.

For the physical system at hand, $\rho_{\rm tot} \equiv \rho_{\rm QED} + \rho_\nu$ and $P_{\rm tot} \equiv P_{\rm QED}+P_\nu$, both of which sum over the QED and the neutrino sectors.%
\footnote{In the time frame of interest, the QED plasma is composed of photons, electrons/positrons, and muons/anti-muons, while the neutrino sector comprises three families of SM neutrinos and anti-neutrinos. We neglect the small, ${\cal O}(10^{-10})$ universal matter--antimatter asymmetry, so that a particle name is taken to refer to both the particle and its anti-particle.}
We assume the QED sector to be always in a state of thermodynamic equilibrium, so that $\rho_{\rm QED}$ and $P_{\rm QED}$ are connected by the standard thermodynamic relation $\rho_{\rm QED} = -P_{\rm QED}+ T \, (\partial /\partial T) P_{\rm QED}$.


\subsection{Boltzmann equation}
\label{sec:Boltzmann}

Schematically, the generalised Boltzmann equation for the one-particle reduced density matrix of the neutrino ensemble, $\varrho(t,p)$, in an FLRW universe is given by
$\partial_t \varrho - p H \partial_p \varrho = - {\rm i} [\mathbb{H},\varrho] + {\cal I}[\varrho]$, where
$[\mathbb{H},\varrho] \equiv \mathbb{H} \varrho-\varrho \mathbb{H}$ denotes a commutator between the flavour oscillations Hamiltonian~$\mathbb{H}$ and~$\varrho$, and the collision integral ${\cal I}[\varrho]$ encapsulates all non-unitary (scattering) effects on $\varrho$~\cite{Sigl:1992fn}.
In terms of the rescaled coordinates, the Boltzmann equation reads~(e.g.,~\cite{Mangano:2005cc})
\begin{align}
\frac{{\rm d}\varrho(x,y)}{{\rm d} x}=
\frac{1}{m_e}\frac{m_e^4}{x^4}\sqrt{\frac{3 m^2_{\rm Pl}}{8\pi\bar{\rho}_{\rm tot} }}
\Bigg(
-{\rm i}
\Big[\mathbb{H} (x,y,z(x)),
\varrho(x,y) \Big]
+{\mathcal{I}}[\varrho(x,y)]
\Bigg),
\label{eq:drho_dx_nxn}
\end{align}
where $m_{\rm Pl}$ is the Planck mass, and we note in passing that the prefactor $ (m_e/x)^2 \! \sqrt{3 m^2_{\rm Pl}/8\pi\bar{\rho}_{\rm tot}}$ is but the inverse Hubble expansion rate $H^{-1}$.

Working in the flavour basis, the oscillations Hamiltonian~\cite{Sigl:1992fn},
\begin{align}
\mathbb{H} (x,y,z(x)) =\frac{x^6}{m_e^6}\frac{U\mathbb{M}U^\dagger}{2y}
-
2\sqrt{2}G_F y
\left(
\frac{{\mathbb{E}}_\ell(z(x))+{\mathbb{P}}_\ell(z(x))}{m_W^2}
+
\frac{4}{3} \frac{{\mathbb{E}}_\nu(x,z(x))}{m_W^2 \cos^2{\theta_W}}
\right),
\label{eq:hamiltonian}
\end{align}
comprises a vacuum and an in-medium part.
The former consists of the neutrino squared-mass difference matrix $\mathbb{M}={\rm diag}(0,\Delta m^2_{21},\Delta m^2_{31})$ and the vacuum mixing matrix $U= R(\theta_{23})R(\theta_{13}) R(\theta_{12})$ parameterised by three Euler rotation angles $\theta_{23}$, $\theta_{13}$, and $\theta_{12}$.
We explicitly set the Dirac $CP$ phase, $\delta_{CP}$, to zero in this analysis.
In this way, there is no neutrino asymmetry, and we can consider
the neutrino momentum distribution equal to the antineutrino one.
We do note, however, that global fits are beginning to indicate a preferred~$\delta_{CP}$ value in the range $[\pi, 2\pi]$~\cite{deSalas:2017kay,deSalas:2020pgw,Esteban:2020cvm,Capozzi:2020qhw}.
The $CP$-conserving value $\delta_{CP}=0$ is excluded at approximately~3$\sigma$, for both neutrino mass orderings.

The in-medium part of the Hamiltonian~\eqref{eq:hamiltonian} contains a $CP$-symmetric (by assumption) correction to the neutrino dispersion relation~\cite{Notzold:1987ik}, i.e., the ``matter potential'', proportional to the Fermi constant $G_F$, with $m_W$ the $W$ boson mass and $\theta_W$ the weak mixing angle. The terms $\mathbb{E}_\ell$ and $\mathbb{P}_\ell$ are momentum-integrals of some combinations of the charged-lepton energy $\epsilon_\ell(y)\equiv (y^2+ a^2 m_\ell^2)^{1/2}$ and (equilibrium) occupation number~$f_\ell(y)$,
\begin{eqnarray}
\mathbb{E}_\ell & \equiv& \frac{1}{\pi^2} \int {\rm d} y \, y^2 \, {\rm diag}\Big(\epsilon_e(y) f_e(y),\epsilon_\mu(y) f_\mu(y),0\Big)
{\underset{\text{ideal gas}}{=}}
{\rm diag}(\bar{\rho}_e,\bar{\rho}_\mu,0),\label{eq:eell}\\
\mathbb{P}_\ell &\equiv& \frac{1}{3 \pi^2} \int {\rm d} y \, y^2 \, {\rm diag}\Bigg(\frac{y^2}{\epsilon_e(y)} f_e(y),\frac{y^2}{\epsilon_\mu(y)} f_\mu(y),0\Bigg)
{\underset{\text{ideal gas}}{=}} {\rm diag}(\bar{P}_e,\bar{P}_\mu,0), \label{eq:pell}
\end{eqnarray}
which coincide respectively with the comoving energy density and pressure of the relevant charged leptons in the ideal gas limit;
$\mathbb{E}_\nu\equiv\pi^{-2}\int \mathrm{d}y\,y^3\varrho(y)$ is the equivalent for an ultra-relativistic neutrino gas.
Note that the $\mathbb{E}_\ell +\mathbb{P}_\ell$ term in equation~\eqref{eq:hamiltonian} differs from its usual presentation found in, e.g., equation~(2.2) of~\cite{Gariazzo:2019gyi}, which has $\mathbb{E}_\ell +\mathbb{P}_\ell$ replaced with $(4/3)\, \mathbb{E}_\ell$.
First reported in~\cite{Notzold:1987ik}, the former is in fact the more general result, while $(4/3)\, \mathbb{E}_\ell$ applies strictly only when the charged leptons are ultra-relativistic. In practice, however, using the incorrect expression incurs an error no larger than $10^{-5}$ in $\Neff^{\rm SM}$.

The collision integral ${\cal I}[\varrho]$ incorporates in principle all weak scattering processes wherein at least one neutrino appears in either the initial or final state.
As in the previous update~\cite{deSalas:2016ztq},
however, we account only for $2 \to 2$ processes involving
(i) two neutrinos and two electrons anyway distributed in the initial and final states, and
(ii) neutrino--neutrino scattering.
Then, schematically,
${\cal I}[\varrho]= {\cal I}_{\nu e}[\varrho] + {\cal I}_{\nu \nu}[\varrho]$ comprises 9D momentum-integrals that, at tree level, can be systematically reduced to 2D integrals of the forms
\begin{eqnarray}
{\cal I}_{\nu e}[\varrho(y)]
&\propto&
G_F^2 \int {\rm d}y_2\, {\rm d}y_3\, \Pi_{\nu e}(y,y_2,y_3;x) \, F_{\nu e}(\varrho(y),\varrho(y_2),f_e(y_3),f_e(y_4)), \label{eq:collision1}
\\
{\cal I}_{\nu \nu}[\varrho(y)]
&\propto&
G_F^2 \int {\rm d}y_2\, {\rm d}y_3\, \Pi_{\nu \nu}(y,y_2,y_3;x) \, F_{\nu \nu}(\varrho(y),\varrho(y_2),\varrho(y_3),\varrho(y_4)),
\label{eq:collision2}
\end{eqnarray}
where $\Pi_{\nu e}$ and $\Pi_{\nu \nu}$ are scalar functions representing the scattering kernels, and the phase space matrices~$F_{\nu e}$ and~$F_{\nu \nu}$ are products of $\varrho(y)$ including quantum statistics.
We defer to section~\ref{sec:collisionintegrals} the discussion of the exact forms and our implementation of ${\cal I}[\varrho]$ in the solution procedure, but note here that this is one of the first studies in which the neutrino--neutrino scattering integral ${\cal I}_{\nu \nu}[\varrho(y)]$ is solved numerically ``as is'' without approximations.

Then, noting that $\bar{\rho}_\nu = {\rm Tr}[\mathbb{E}_\nu]$ and correspondingly $\bar{P}_\nu = \bar{\rho}_\nu/3$,
equations~\eqref{comoving_conserv_equation} to \eqref{eq:collision2} form a closed system which we shall solve numerically using the precision neutrino decoupling code~\fortepiano~\cite{Gariazzo:2019gyi}.


\section{New elements in this calculation}
\label{sec:whatsnew}

Aside from a new code, three aspects of our new computation of $\Neff^{\rm SM}$ sets it apart from most existing calculations.
Firstly, new and significant finite-temperature QED corrections to ${\cal O}(e^3)$  identified in~\cite{Bennett:2019ewm} are incorporated into the precision calculation, including an ${\cal O}(e^2)$ logarithmic correction that has never been implemented in previous calculations and whose impact we demonstrate in this work to be below the current numerical error. Secondly, we solve numerically the full neutrino--neutrino collision integral --- without approximation --- in the presence of neutrino oscillations, as was previously attempted only in reference~\cite{Froustey:2020mcq}.
Thirdly, we identify and quantify the various sources of errors that can impact on the accuracy with which the benchmark $\Neff^{\rm SM}$ can be computed theoretically; our treatment in this regard is similar to that  of reference~\cite{Froustey:2020mcq}, but describes more extensively the impact of neutrino oscillation parameters beyond the constraints from oscillation experiments.
We discuss these three new elements in three subsections below.


\subsection{New finite-temperature QED corrections}
\label{system:QED}

Finite-temperature QED affects $\Neff^{\rm SM}$ in two distinct ways~\cite{Bennett:2019ewm}. The dominant effect is an altered equation of state for the QED plasma, through which the bulk thermodynamic quantities of the plasma --- such as $\rho_{\rm QED}$ and $P_{\rm QED}$ that appear in the continuity equation~\eqref{comoving_conserv_equation} --- are modified away from their ideal gas predictions in a temperature-dependent manner.
A secondary effect are temperature-dependent corrections to the scattering rates of the constituent QED (quasi)particles with neutrinos; these enter at a practical level into
the neutrino Boltzmann equation~\eqref{eq:drho_dx_nxn} through the ${\cal I}_{\nu e}[\varrho]$ collision integral~\eqref{eq:collision1} and, albeit minutely, through the in-medium neutrino dispersion relation in the flavour oscillation Hamiltonian~\eqref{eq:hamiltonian}.


\subsubsection{Finite-temperature corrections to the QED equation of state}
\label{sec:ftqedeos}

\begin{figure}[t]
\centering
\includegraphics[width=14cm]{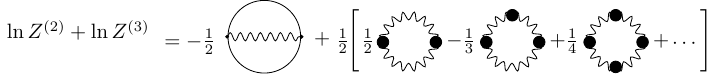}
\caption{Diagrammatic expression for the ${\cal O}(e^2)$ and ${\cal O}(e^3)$ corrections to the finite-temperature QED partition function.
The filled black circles represent one-particle irreducible photon self-energies at finite temperature.}
\label{partition_function_diagrams}
\end{figure}

The finite-temperature QED partition function $Z$ is well established in the literature~(see, e.g.,~\cite{Kapusta:2006pm}), to a level of precision (as measured by powers in the elementary electric charge $e$) higher than is necessary for the present calculation~\cite{Bennett:2019ewm}. The diagrammatic representation of the ${\cal O}(e^2)$ and ${\cal O}(e^3)$ corrections to $\ln Z$ used in this work are shown in figure~\ref{partition_function_diagrams}. Contributions at higher orders in $e$ have been estimated in the instantaneous decoupling limit to contribute at most $\delta \Neff \sim 4 \times 10^{-6}$~\cite{Bennett:2019ewm} --- substantially below the intrinsic numerical noise of our computations (see section~\ref{sec:convergence}) ---
and are thus neglected in the benchmark calculation.


The leading-order, $\mathcal{O}(e^2)$ correction contributes $\delta \Neff \sim 0.01$~\cite{Dicus:1982bz,Heckler:1994tv,Lopez:1998vk,Bennett:2019ewm}, and is often discussed in two parts:
a logarithmic contribution and a log-independent one (see equation~(4.7) of~\cite{Bennett:2019ewm}). The latter contribution
modifies the continuity equation~\eqref{comoving_conserv_equation} as described in~\cite{Mangano:2001iu} (see also appendix B of~\cite{Gariazzo:2019gyi}), and is now a staple ingredient in the computation of $\Neff^{\rm SM}$~\cite{Mangano:2005cc,Birrell:2014uka,Grohs2016,Escudero:2018mvt,Escudero:2020dfa,Froustey:2019owm,deSalas:2016ztq,Akita:2020szl,Gariazzo:2019gyi}.
The former, on the other hand, has been hitherto regularly neglected, but should contribute a borderline-significant $\delta \Neff \sim -5 \times 10^{-5}$~\cite{Bennett:2019ewm}. In this work, we include for the first time this $\mathcal{O}(e^2)$ logarithmic contribution in the precision computation of $\Neff^{\rm SM}$, through further modifications to the continuity equation~\eqref{comoving_conserv_equation} following~\cite{Bennett:2019ewm}.

Of greater interest, however, is the new, sub-leading $\mathcal{O}(e^3)$ correction arising from the resummation of ring diagrams depicted enclosed in square brackets in figure~\ref{partition_function_diagrams}. Reference~\cite{Bennett:2019ewm} estimates a sizeable contribution from this set of diagrams: $\delta\Neff\sim-10^{-3}$, larger than the effect of including neutrino flavour oscillations. Following the procedure laid down in~\cite{Bennett:2019ewm}, the said correction was recently adopted in~\cite{Akita:2020szl,Froustey:2020mcq} in their computation of $\Neff^{\rm SM}$. We likewise incorporate this $\mathcal{O}(e^3)$ correction into our present precision calculation.


\subsubsection{Finite-temperature QED corrections to weak scattering rates}
\label{sec:weakrates}

\begin{figure}[t]
\centering
\includegraphics[width=15cm]{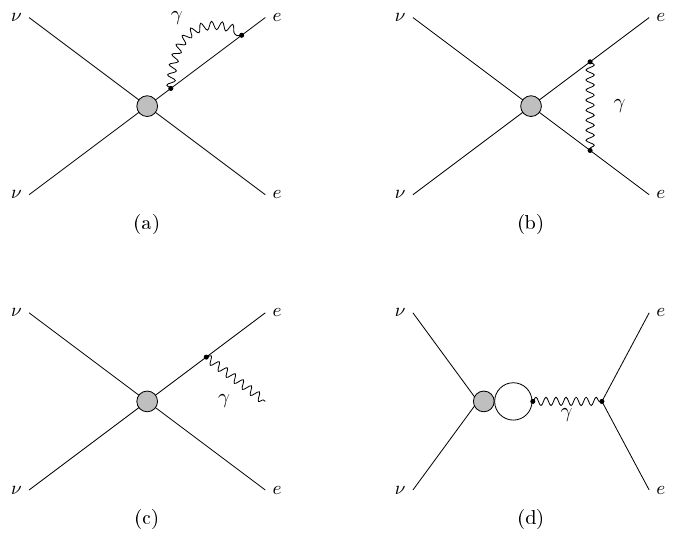}
\caption{The only four qualitatively different finite-temperature QED corrections to the weak scattering rates.}
\label{QED2to2_scatterings}
\end{figure}

We distinguish four types of finite-temperature QED corrections to the 4-Fermi contact interaction that describes $2 \to2$ neutrino--electron scattering at leading order: (a)~modification to the dispersion relation,
(b)~vertex corrections, (c)~real emission or absorption, and (d)~closed fermion loops~\cite{Tomalak:2019ibg,Hill:2019xqk}. These are depicted diagrammatically in figure~\ref{QED2to2_scatterings}.

Contributions of the type (a) amount to dressing the fermionic QED-charged propagator with a photon~\cite{Silin:1960pya}, which, in the quasiparticle approximation, can be effectively captured by way of a modified, in-medium dispersion relation in the evaluation of the relevant Feynman diagrams.
In practice, diagrams of this sort are often evaluated using partially resummed propagators, wherein the (vacuum) particle mass is shifted to its thermal counterpart
obtained from the self-energy computed to some fixed order in $e$ --- hence the common name ``thermal mass correction''.%
\footnote{The procedure of shifting the pole mass of the thermal propagator to account for finite-temperature corrections is generally valid at the {\it diagrammatic} level within the so-called quasiparticle approximation ~\cite{Arnold:2002zm,Anisimov:2008dz,Drewes:2010pf}. It is {\it not} a valid procedure, however, to compute finite-temperature corrections to bulk thermodynamic quantities (e.g., energy density, pressure, etc.)~by replacing the vacuum particle mass with its thermal counterpart in expressions that have been established {\it originally for an ideal gas}.
See~\cite{Bennett:2019ewm} for a discussion of this issue in relation to the $\Neff^{\rm SM}$ computation.}
In doing so, we actually resum infinitely many QED self-interactions, but at the same time also neglect the subdominant effects of additional quasiparticle solutions due to collective excitations in the plasma~\cite{Klimov:1982bv,Weldon:1982bn}.

In practice, implementation of the above description amounts to replacing all occurrences of $m_e^2$ in the 9D weak collision integral ${\cal I}[\varrho]$ with its thermal counterpart $m_e^2+\delta m_e^2(p,T)$, where we take $\delta m_e^2(p,T)$ to be the ${\cal O} (e^2)$, one-loop electron self-energy given in, e.g., equation~(4.38) of~\cite{Bennett:2019ewm}. This is an especially trivial task if we ignore the momentum-dependent piece in
the said equation --- which has been shown to contribute less than 10\% to the total $\delta m_e^2(p,T)$~\cite{Mangano:2001iu} --- so
that the reduction of ${\cal I}[\varrho]$ to 2D proceeds identically as in the case with bare propagators.
The same procedure can also be applied to equations~\eqref{eq:eell} and~\eqref{eq:pell} to approximately correct the matter potentials from the charged-lepton background.

Such a simple implementation has previously been adopted for ${\cal I}[\varrho]$ in the precision calculations of~\cite{deSalas:2016ztq,Gariazzo:2019gyi}, as well as in the estimates of~\cite{Bennett:2019ewm} which found this correction to contribute $\delta \Neff \sim - 5 \times 10^{-5}$. For simplicity and in view of the size of the correction, we shall adopt the same approach in this work and include in addition a similarly-corrected matter potential. We stress however that the procedure is, strictly speaking, inconsistent, and leave a careful treatment of thermal-mass correction to the weak rates for future work.

Contributions of the types (b), (c) and (d) unfortunately cannot be dealt with in a similar fashion. An explicit evaluation of each individual diagram is required to assess their significance --- a tedious task (see, e.g.,~\cite{Tomalak:2019ibg,Hill:2019xqk} for the zero-temperature case) that we also leave for a future publication.


\subsection{Full collision integral for neutrino--neutrino scattering}
\label{sec:full}

As discussed in section~\ref{sec:Boltzmann}, we split the weak collision integral ${\cal I}[\varrho]$ into two parts, ${\cal I}_{\nu e}[\varrho]$ and ${\cal I}_{\nu \nu}[\varrho]$, accounting respectively for $2 \to 2$ neutrino--electron and neutrino--neutrino scattering processes.
In the flavour basis and assuming the neutrino density matrix $\varrho(y)$ to be diagonal,
the {\it diagonal} elements of both ${\cal I}_{\nu e}[\varrho]$ and ${\cal I}_{\nu \nu}[\varrho]$, $\left\{ {\mathcal{I}}[\varrho(y)]\right\}_{\alpha \alpha}$, where $\alpha=e,\mu,\tau$ is a flavour index, are identically the standard Boltzmann collision integral due to the said $2 \to 2$ scattering processes
for the occupation number of $\nu_\alpha$ at mode $y$.
The {\it off-diagonal} elements, $\left\{ {\mathcal{I}}[\varrho(y)]\right\}_{\alpha \beta}$, where $\alpha \neq \beta$, are on the other hand peculiar to flavour oscillations and responsible for quantum decoherence of the neutrino ensemble.

The general, 9D forms of ${\cal I}_{\nu e}[\varrho]$ and ${\cal I}_{\nu \nu}[\varrho]$ have been determined in~\cite{Sigl:1992fn,McKellar:1992ja}.
Practical implementation, however, requires that we first reduce them to 2D integrals of the forms~\eqref{eq:collision1} and~\eqref{eq:collision2}, wherein each integrand comprises a scattering kernel $\Pi$ and a phase space matrix~$F$. Many integral reduction procedures exist for the purpose (e.g.,~\cite{Dolgov:1997mb,Hannestad:2015tea,Bennett:2019ewm}), differing from each other only in the order in which the angular dependences of the integrand are eliminated in the intermediate steps.
For ${\cal I}_{\nu e}[\varrho]$ we use the expressions presented in appendix A of~\cite{deSalas:2016ztq}, the same expressions that have been hard-coded in \fortepiano\ for a previous study~\cite{Gariazzo:2019gyi}. For completeness, we present them again in appendix~\ref{sec:collisionintegralsapp} of this work.

On the other hand, the ${\cal I}_{\nu \nu}[\varrho]$ collision integral~\eqref{eq:collision2} is highly nonlinear: its phase space matrix~$F_{\nu \nu}$ couples $\varrho(y)$ at four different modes. For this reason, with the exception of the recent study of~\cite{Froustey:2020mcq} which does consider the integral in full, ${\cal I}_{\nu \nu}[\varrho]$ has thus far mostly been treated only in an approximate fashion.
The 2016 update of~\cite{deSalas:2016ztq}, for example, replaced its diagonal entries with the standard Boltzmann collision integrals for the occupation numbers --- a procedure equivalent to assuming a flavour-diagonal~$\varrho(y)$ in the evaluation of~$F_{\nu \nu}$ ---
while modelling the off-diagonal elements using a damping approximation (to be discussed in subsection~\ref{sec:collisionintegrals}). Reference~\cite{Gariazzo:2019gyi} likewise adopted the off-diagonal damping approximation, but opted to neglect the diagonal entries. Meanwhile, the recent study of~\cite{Akita:2020szl}, while treating the diagonal entries in the manner of~\cite{deSalas:2016ztq}, did not model the off-diagonal ones at all.

In the present work, we incorporate the full ${\cal I}_{\nu \nu}[\varrho]$ collision integral ``as is'' in the precision computation of the benchmark $\Neff^{\rm SM}$. The complete expression of its 2D reduced form can be found in appendix~\ref{sec:collisionintegralsapp}.
Because it couples $\varrho(y)$ at four different modes~$y$, numerical evaluation of the phase space matrix $F_{\nu \nu}$ always requires a minimum of one interpolation of~$\varrho(y)$ in $y$-space in order to enforce energy conservation amongst the four neutrino states participating in a $\nu \nu \to \nu \nu$ collision process. Linear interpolation between two nodes closest to the desired $y$-value suffices for the off-diagonal entries $\{\varrho(y)\}_{\alpha \beta}$, $\alpha \neq \beta$. For the diagonal entries, $\{\varrho(y)\}_{\alpha \alpha}$, we find that interpolating the normalised $\{\varrho(y) \}_{\alpha \alpha}/f_{\rm eq}(y)$, where $f_{\rm eq}$ is the relativistic Fermi--Dirac distribution, generally offers better stability than interpolating the bare $\{\varrho(y) \}_{\alpha \alpha}$. The most critical numerical issue in this regard, however, is the choice of momentum discretisation scheme. We shall discuss this in detail in section~\ref{sec:convergence}.


\subsection{Assessment of remaining uncertainties}
\label{sec:uncertainties}

Aside from the ${\cal O}(e^4)$ and higher finite-temperature corrections to the QED equation of state that have been previously quantified in~\cite{Bennett:2019ewm}, uncertainties in the SM benchmark $\Neff^{\rm SM}$ can arise also from our treatment of the weak sector. To this end, we identify three classes of errors: (i)~optional physical approximations provided in~\fortepiano\ to stabilise and expedite the evaluation of the weak collision integrals,
(ii)~measurement errors in the physical parameters of the neutrino sector, and (iii)~numerical non-convergence originating in the discretisation and initialisation procedures of the solution scheme. The last class of errors, class~(iii), is inherent in all numerical approximations to solutions of differential equations, and warrants a separate discussion in section~\ref{sec:convergence}.
Error sources~(i) and~(ii) are physical in origin, which we describe in two subsections below.


\subsubsection{Approximate treatments of the weak collision integrals}
\label{sec:collisionintegrals}

Since tracking the decoupling of the neutrino sector from the QED plasma is the main objective of our calculation, the minimum set-up of~\fortepiano\ always includes a full and non-negotiable numerical evaluation of the diagonal components of the neutrino--electron collision integral~${\cal I}_{\nu e}[\varrho]$. Beyond this bare minimum requirement, however, several approximations and/or alternative implementations of the remaining ${\cal I}[\varrho]$ terms may be considered to facilitate their computation. Reference~\cite{Gariazzo:2019gyi}, for example, assumed $\{{\cal I}_{\nu \nu}[\varrho]\}_{\alpha \alpha}=0$ and a damping approximation for all off-diagonal entries in what we shall call the {\bf minimum set-up}.

\paragraph{Diagonal neutrino--neutrino collision integral.}
There are good reasons to think that the diagonal entries of the neutrino--neutrino collision integral, $\{{\cal I}_{\nu \nu}[\varrho]\}_{\alpha \alpha}$,
may be effectively dispensable. Phenomenologically, the $\{{\cal I}_{\nu \nu}[\varrho]\}_{\alpha \alpha}$ terms serve to transport energy between different neutrino flavours, a role that may conceivably be fulfilled to a good extent by large-mixing flavour oscillations in those studies that account for the latter. In view that ${\cal I}_{\nu \nu}[\varrho]$ is highly nonlinear and its full evaluation comes with substantial numerical uncertainties (see section~\ref{sec:fullnunu}), alternative implementations of~${\cal I}_{\nu \nu}[\varrho]$ may be desirable for a numerically stable outcome, {\it provided} of course that accuracy is not compromised.

In this work, we test three different implementations of $\{{\cal I}_{\nu \nu}[\varrho]\}_{\alpha \alpha}$:
\begin{enumerate}
\item $\{{\cal I}_{\nu \nu}[\varrho]\}_{\alpha \alpha}=0$, as in~\cite{Gariazzo:2019gyi}: the minimum set-up uses this approximation;
\item As in~\cite{deSalas:2016ztq,Akita:2020szl}, we evaluate $\{{\cal I}_{\nu \nu}[\varrho]\}_{\alpha \alpha}$ assuming a {\bf diagonal} $\varrho(y)$; and
\item A {\bf full} evaluation of $\{{\cal I}_{\nu \nu}[\varrho]\}_{\alpha \alpha}$ without assumptions;
\end{enumerate}
and we reiterate that the last named option is itself a novel element of the present work. See section~\ref{sec:full}.

\paragraph{Off-diagonal damping approximation.}
If the deviation of the neutrino ensemble from equilibrium is minimal, the off-diagonal entries of the weak collision integral, $\{{\cal I}[\varrho(y)]\}_{\alpha \beta}$, where $\alpha \neq \beta$, can be systematically manipulated into the form
\begin{equation}
\big\{ {\mathcal{I}}[\varrho(y)]\big\}_{\alpha \beta} = - \big\{D(x,y,z) \big\}_{\alpha \beta} \big\{ \varrho(y) \big\}_{\alpha \beta}
\label{eq:damping},
\end{equation}
where $\{D(x,y,z)\}_{\alpha \beta} =\{D_{\nu e}(x,y,z)\}_{\alpha \beta} + \{D_{\nu \nu}(x,y,z) \}_{\alpha \beta}$ are flavour-dependent damping coefficients. Equation~\eqref{eq:damping} is the so-called ``damping approximation''. At the practical level, the approximation effectively decouples the evolution of $\{ \varrho(y)\}_{\alpha \beta}$ not only from other momentum modes $y'\neq y$ but also from other entries of $ \varrho(y)$ at the same mode~$y$, and thus, where it is valid, offers an extremely efficient computational pathway.%
\footnote{There is in principle also a similar damping approximation for the diagonal entries of ${\cal I}[\varrho]$. See appendix~\ref{sec:dampingcoefficients}. We do not however use it in our computation of $\Neff^{\rm SM}$.}

In the context of precision $\Neff^{\rm SM}$ computation and related topics (e.g., sterile neutrino thermalisation), there are historically two understandings of the damping approximation~\eqref{eq:damping} and hence two disparate sets of damping coefficients in the existing literature derived under different assumptions:
\begin{enumerate}
\item At each mode $y$, standard linear response instructs us to equate all occurrences of $\varrho(y')$ in the phase space matrix~$F$ --- {\it except} for $y'=y$ --- to their equilibrium expectations, i.e., flavour-diagonal and $\{ \varrho(y)\}_{\alpha \alpha}=f_{\rm eq}(y)$, where $f_{\rm eq}(y)$ is some equilibrium distribution.
The procedure immediately renders~$\{{\mathcal{I}}[\varrho(y)]\}_{\alpha \beta}$ into the form~\eqref{eq:damping}, where any remaining integration can be immediately performed to yield the damping coefficients.

This approach was previously used in~\cite{Bell:1998ds} to compute $\{D(x,y,z)\}_{\alpha \beta}$ in the $m_e=0$ (i.e., $x=0$) limit assuming Maxwell--Boltzmann statistics. In this work, we generalise these $x=0$
calculations to Fermi--Dirac statistics including Pauli blocking, and find
\begin{eqnarray}
\big\{D_{\nu \nu}(y,z) \big\}_{\alpha \beta}
\! &=& \! {\cal D}(y,z) \equiv \frac{ 2\, G_F^2 \, y\, z^4}{(2 \pi)^3} d(y/z),
\label{eq:nunu} \\
\big\{D_{\nu e}(y,z) \big\}_{\alpha \beta}
\! &=& \! \frac{1}{ 8} \left[(2 \sin^2 \theta_W \pm 1)^2_\alpha + (2 \sin^2 \theta_W \pm 1)^2_\beta+ 8 \sin^4 \theta_W\right] {\cal D}(y,z),
\label{eq:nue}
\end{eqnarray}
where the notation $(2 \sin^2 \theta_W\pm 1)_\alpha$ indicates that the ``$+$'' sign applies to $\alpha =e$ and ``$-$'' to $\alpha = \mu, \tau$. The function $d(s=y/z)$ is a number of order 100 shown in figure~\eqref{eq:dy}.
Its exact form --- expressed as a double momentum integral --- is given in equation~\eqref{eq:selfdamping}; for computational ease, however, we also supply a fitting function $d_{\rm fit}(s)$ in equation~\eqref{eq:fittingfunction}, which reproduces $d(s)$ in the interval $s \in [10^{-4},10^3]$ to better than 0.25\% accuracy.
Details of the calculation can be found in appendix~\ref{sec:dampingcoefficients}.

\item Reference~\cite{McKellar:1992ja} proposed to simplify $\{{\cal I}[\varrho(y)]\}_{\alpha \beta}$ under the ansatz
\begin{equation}
\varrho(y) = \frac{f_{\rm eq}(y)}{f_{\rm eq}(\langle y \rangle)}\, \varrho(\langle y \rangle),
\label{eq:mtansatz}
\end{equation}
such that $\varrho(y)$ at all modes~$y$ evolve {\it in phase}, where $\langle y \rangle$ denotes a representative momentum. Upon integration in $y$, the procedure yields a thermally-averaged $\langle \{{\cal I}[\varrho(y)]\}_{\alpha \beta}\rangle$, expressed in the damping form~\eqref{eq:damping} in terms of $\varrho(\langle y \rangle)$ and
a set of thermally-averaged damping coefficients
$\langle \{D(x,y,z)\}_{\alpha \beta} \rangle$.
In introducing the ansatz~\eqref{eq:mtansatz}, the primary motivation of~\cite{McKellar:1992ja} was to reduce the generalised Boltzmann equation~\eqref{eq:drho_dx_nxn} to a single set of ``quantum rate equations'' evaluated at the representative momentum~$\langle y \rangle$. We note however that some subsequent works~(e.g., \cite{Kainulainen:2001cb}) invoked a scaling argument to obtain $y$-dependent damping coefficients such as $\{D(x,y,z)\}_{\alpha \beta} = [y/\langle y \rangle] \langle \{D(x,y,z)\}_{\alpha \beta} \rangle$.

Physically, the ansatz~\eqref{eq:mtansatz} effectively removes those parts of the collision integral~${\cal I}[\varrho]$ associated with flavour-blind elastic scattering. The $y$-scaled versions of the damping coefficients derived under this scheme, particularly for~${\cal I}_{\nu \nu}[\varrho]$, have historically been employed in the precision $\Neff^{\rm SM}$ calculations of~\cite{deSalas:2016ztq,Gariazzo:2019gyi}. We do not consider them in this work, however, as our goals differ substantively from the original intentions of~\cite{McKellar:1992ja} that motivated the ansatz~\eqref{eq:mtansatz}.%
\footnote{One could also interpret the ansatz~\eqref{eq:mtansatz} physically as an assumption of a time-scale hierarchy between flavour-blind elastic scattering and other (inelastic and flavour-dependent) processes, whereby the former are taken to be always fast enough to set all $\varrho(y)$ in phase, while the latter determine the slower aspects of $\varrho(y)$'s evolution. For ultrarelativistic neutrinos that only have weak interactions and hence only one collision time-scale, the assumption of such a hierarchy is unfortunately ill-conceived. The ``in phase'' assumption is likewise not borne out by our numerical results. These constitute more reasons to reject the ansatz~\eqref{eq:mtansatz}.}

\end{enumerate}

In the present study, we test two different implementations of $\{ {\mathcal{I}}[\varrho(y)]\}_{\alpha \beta}$:
(i)~a full, real-time numerical evaluation of all entries, and
(ii)~$\{ \mathcal{I}_{\nu e}[\varrho(y)]\}_{\alpha \beta}$ and $\{ \mathcal{I}_{\nu \nu}[\varrho(y)]\}_{\alpha \beta}$ computed under the damping approximation~\eqref{eq:damping}, with damping coefficients~\eqref{eq:nunu} and~\eqref{eq:nue} obtained from linear response.
The latter implementation corresponds to the minimum set-up.


\subsubsection{Measurement errors in the physical parameters of the neutrino sector}

The fine structure constant $\alpha \equiv e^2/4 \pi$ and the electron mass $m_e$ have both been experimentally determined to nine significant digits~\cite{codata} (see table~\ref{tab:uncertainties}). In the context of computing finite-temperature corrections to $\Neff^{\rm SM}$, these parameters are essentially infinitely well known.

\begin{table}[t]
	\centering
	\begin{tabular}{|l|c|c|c|c|}
		\hline
		&Parameter [Units] &Value $\pm$ 1$\sigma$ uncertainty &Reference \\
		\hline
		\parbox[t]{4mm}{\multirow{2}{*}{\rotatebox[origin=c]{90}{QED}}}
		&$\alpha/10^{-3}$ & $7.297 352 5693 \pm 0.0000000011 $ & \cite{codata} \\
		&$m_e\, [{\rm MeV}]$& $0.510 998 950 00 \pm 0.00000000015$ & \cite{codata} \\
		\hline
		\parbox[t]{4mm}{\multirow{6}{*}{\rotatebox[origin=c]{90}{Weak}}}
		&$\sin^2 \theta_W$ & $0.23871 \pm 0.00009$ & \cite{Kumar:2013yoa,Erler:2013xha} \\
		& $g_L$& $0.727$ & \cite{Kumar:2013yoa,Erler:2013xha} \\
		& $\tilde g_L$& $-0.273$ & \cite{Kumar:2013yoa,Erler:2013xha} \\
		& $g_R$& $0.233$ & \cite{Kumar:2013yoa,Erler:2013xha} \\
		&$G_F\, [10^{-5} {\rm GeV}^{-2}]$& $1.166 378 7 \pm 0.0000006$ & \cite{Zyla:2020zbs} \\
		& $m_W\, [{\rm GeV}]$& $80.379 \pm 0.012$ & \cite{Zyla:2020zbs} \\
		\hline
		\parbox[t]{4mm}{\multirow{6}{*}{\rotatebox[origin=c]{90}{Neutrino}}}
		& $\sin^2 \theta_{12}/10^{-1}$& $3.18 \pm 0.16$ & \cite{deSalas:2020pgw} \\
		& $\sin^2 \theta_{13}/10^{-2}$& $2.200^{+0.069}_{-0.062}$& \cite{deSalas:2020pgw} \\
		&$\sin^2 \theta_{23}/10^{-1}$ &$5.74 \pm 0.14$ & \cite{deSalas:2020pgw} \\
		&$\Delta m^2_{21}\, [10^{-5} {\rm eV}^2]$ &$7.50^{+0.22}_{-0.20}$ & \cite{deSalas:2020pgw} \\
		&$\Delta m^2_{31}\, [10^{-3} {\rm eV}^2]$ (NO) &$2.55^{+0.02}_{-0.03}$ & \cite{deSalas:2020pgw} \\
		&$\Delta m^2_{31}\, [10^{-3} {\rm eV}^2]$ (IO) &$-2.45^{+0.03}_{-0.02}$ & \cite{deSalas:2020pgw} \\
		\hline
	\end{tabular}
	\caption{Central values and $1\sigma$ uncertainties of the physical constants used in this work. The neutrino parameter values have been derived in the global fit~\cite{deSalas:2020pgw} assuming a normal mass ordering (NO), with the exception of the last entry which assumes an inverted mass ordering~(IO).
	The weak mixing angle value quoted here corresponds to
	$\sin^2 \theta_W (0)_{\overline{\rm MS}}$ in the modified minimal subtraction ($\overline{\rm MS}$) scheme, established via SM renormalisation group running from the measured value of $\sin^2 \theta_W (m_Z)_{\overline{\rm MS}} = 0.23124 \pm 0.00006$ at the $Z$-pole~\cite{Kumar:2013yoa,Erler:2013xha}.}
	\label{tab:uncertainties}
\end{table}

Likewise, the small uncertainties in the weak sector constants --- $G_F$, $\sin^2 \theta_W$, $m_W$, $g_L$, $\tilde{g}_L$, and $g_R$ ---
are not expected to impact on $\Neff^{\rm SM}$ by more than $|\delta \Neff|\sim 10^{-5}$, the intrinsic numerical noise of the computation (see section~\ref{sec:convergence}).
To see this, note that as a rule of thumb, shifting the neutrino decoupling temperature $T_d$ by a fractional $\sim 0.1\%$ induces a change of $|\delta \Neff| \lesssim 10^{-4}$~\cite{Bennett:2019ewm}.
Then, by simple power counting, varying $G_F$ even by as much as $3\sigma$ away from its central values
(i.e., relative change of $\sim \pm 10^{-4}\%$) can only have a negligible effect on~$\Neff^{\rm SM}$.
The impact of varying the $W$-boson mass $m_W$ must be similarly imperceptible, since in our parameterisation $m_W$ enters the picture only through flavour oscillations,
which are themselves known to be a subdominant, at most $|\delta \Neff|\sim 0.001$ effect. For $\sin^2 \theta_W$, we have explicitly tested that substituting $\sin^2 \theta_W (m_Z)_{\overline{\rm MS}}$ for $\sin^2 \theta_W (0)_{\overline{\rm MS}}$ (see caption of table~\ref{tab:uncertainties} for definitions) shifts $\Neff^{\rm SM}$ by a minute $|\delta \Neff|\sim 4 \times 10^{-5}$.

Physical parameters of the neutrino sector are, on the other hand, generally far less well measured. The most poorly-determined parameter from the global fit of~\cite{deSalas:2020pgw} for example, the solar neutrino mixing angle $\sin^2 \theta_{12}$, has a $1 \sigma$-uncertainty of about 5\%, as shown in table~\ref{tab:uncertainties}.
This is
not to mention the as-yet-undetermined sign of $\Delta m^2_{31}$, i.e., a normal or inverted ordering of the neutrino masses, although global fits tend to favour $\Delta m^2_{31}>0$ (normal ordering) marginally~\cite{deSalas:2017kay, Esteban:2018azc, deSalas:2018bym, deSalas:2020pgw,Esteban:2020cvm,Capozzi:2020qhw}.

In this work, we compute our SM benchmark $\Neff^{\rm SM}$ using the central measured values of the physical parameters given in table~\ref{tab:uncertainties}, assuming a normal ordering of the neutrino masses ($m_3 > m_2 > m_1$).
To explore the dependence of $\Neff^{\rm SM}$ on the parameters of the neutrino sector,
we vary the neutrino mass splittings and vacuum mixing angles one at a time by well beyond $5 \sigma$ away on both sides from their central values.
Such a large range of variations more than covers the small differences in the numbers obtained from the three independent global fits of~\cite{deSalas:2020pgw,Esteban:2020cvm,Capozzi:2020qhw}.
For completeness, we repeat the exercise for an inverted mass ordering ($m_2 > m_1 > m_3$), assuming the same parameter values and uncertainties given in table~\ref{tab:uncertainties}.%
\footnote{The global best-fit values vary slightly between our choice of neutrino mass ordering~\cite{deSalas:2020pgw,Esteban:2020cvm,Capozzi:2020qhw}, although the variations are statistically insignificant ($\lesssim1 \sigma$).}


\section{Numerical convergence}
\label{sec:convergence}

We compute $\Neff^{\rm SM}$ using the neutrino decoupling code~\fortepiano, which employs the {\tt DLSODA} routine from the {\tt ODEPACK} Fortran package~\cite{odepack1,odepack2}
to propagate the Boltzmann--continuity system of equations.
Before presenting our results in section~\ref{sec:results}, we perform first in this section a series of numerical convergence tests to ensure that the settings in \fortepiano\ satisfy detailed balance at a level adequate for the computation of $\Neff^{\rm SM}$ to the
desired precision.

To begin with, we note that for an absolute and relative tolerance in {\tt DLSODA} fixed at~$10^{-7}$, the final outcome $\Neff$ may differ by $|\delta \Neff| \sim 10^{-5}$ depending on our choices of compiler ({\tt Intel Fortran} or {\tt GFortran}) and even computer on which to run the code. This latter number therefore forms our estimate of the intrinsic numerical noise of the $\Neff^{\rm SM}$ calculation, and sets a baseline for the assessment of numerical convergence.


\subsection{Convergence variables}
\label{sec:convergencevariables}

Two more settings at our disposal in the execution of \fortepiano\ may impact on the numerical convergence of the final SM benchmark $\Neff^{\rm SM}$: (i)~our choice of momentum binning, both in terms of the binning method and the number of bins used, and (ii)~the initialisation time~$x_{\rm in}$. 
We elaborate on our choices in these regards below.


\subsubsection{Momentum discretisation}
\label{sec:momentum}

To numerically solve the Boltzmann equation~\eqref{eq:drho_dx_nxn},
\fortepiano\ discretises the neutrino momentum distribution on a $y$-grid.
Several discretisation schemes and quadrature methods exist for the purpose. The original version of \fortepiano\ employed in reference~\cite{Gariazzo:2019gyi} to study light sterile neutrino thermalisation uses the method of Gauss--Laguerre (GL) quadrature, which proves to be very efficient at sampling spectral distortions at low momenta, reducing significantly discretisation-related numerical errors at a smaller computational cost.

\begin{table}[t]
	\centering
	\begin{tabular}{|c|c|c||c|c|}
		\hline
		$N_y$ grid
		& $\left\{{\cal I}_{\nu\nu}[\varrho]\right\}_{\alpha \alpha}$
		& $N_y$
		& $\Neff^{\rm SM}$ (no osc)
		& $\Neff^{\rm SM}$ (NO) \\
		\hline
		\hline
		\multirow{9}{*}{\rotatebox[origin=c]{0}{GL}}
		&\multirow{3}{*}{$\left\{{\cal I}_{\nu\nu}[\varrho]\right\}_{\alpha \alpha}=0$}
		&  40 & 3.0426 & 3.0436\\
		&& 60 & 3.0426 & 3.0435\\
		&& 80 & 3.0425 & 3.0435\\
		\cline{2-5}
		&\multirow{3}{*}{\rotatebox[origin=c]{0}{Diagonal $\varrho$}}
		&  40 & 3.0434 & 3.0442\\
		&& 60 & 3.0433 & 3.0441\\
		&& 80 & 3.0433 & 3.0441\\
		\cline{3-5}
		&\multirow{3}{*}{\rotatebox[origin=c]{0}{Full}}
		&  40 & 3.0434 & 3.0439\\
		&& 60 & 3.0433 & 3.0439\\
		&& 80 & 3.0433 & 3.0439\\
		\hline
		\multirow{9}{*}{\rotatebox[origin=c]{0}{NC}}
		&\multirow{3}{*}{$\left\{{\cal I}_{\nu\nu}[\varrho]\right\}_{\alpha \alpha}=0$}
		&  40 & 3.0428 & 3.0438\\
		&& 60 & 3.0426 & 3.0436\\
		&& 80 & 3.0426 & 3.0436\\
		\cline{2-5}
		&\multirow{3}{*}{\rotatebox[origin=c]{0}{Diagonal $\varrho$}}
		&  40 & 3.0436 & 3.0444\\
		&& 60 & 3.0434 & 3.0442\\
		&& 80 & 3.0433 & 3.0442\\
		\cline{3-5}
		&\multirow{3}{*}{\rotatebox[origin=c]{0}{Full}}
		&  40 & 3.0436 & 3.0441\\
		&& 60 & 3.0434 & 3.0439\\
		&& 80 & 3.0433 & 3.0439\\
		\hline
	\end{tabular}
	\caption{Convergence of $\Neff^{\rm SM}$ under variations of the momentum discretisation scheme (Gauss--Laguerre vs Newton--Cotes) and the number of $y$-nodes $N_y$, for three different implementations of the diagonal entries of the neutrino--neutrino collision integral, $\{ {\cal I}_{\nu \nu} [\varrho]\}_{\alpha \alpha}$. See section~\ref{sec:collisionintegrals} for an explanation of the implementations.
	We report results with no neutrino oscillations (no osc) and with oscillations assuming a normal mass ordering (NO) and the central neutrino mixing parameter values of table~\ref{tab:uncertainties}.}
	\label{tab:converge1}
\end{table}

In this work, in addition to GL quadrature, we explore also the method of Newton--Cotes quadrature. We briefly describe these two quadrature methods below and comment on the convergence properties of $\Neff$ under these different discretisation schemes.

\paragraph{Gauss--Laguerre (GL) quadrature}
The method of generalised GL quadrature estimates an integral of the type $\int_0^\infty {\rm d} y \, y^n e^{-y} f(y)$ using $N_y$ weighted nodes on the $y$-grid corresponding to the first $N_y$ roots, $y_i$, of the Laguerre polynomial of some order $N$.
Observe that we do not use all $N$ roots, both because of computational ease and because we expect the higher $y$-nodes' contributions to $\Neff$ to be exponentially suppressed.
Rather, given a user-specified~$y_{\rm max}$,
\fortepiano\ automatically selects only those nodes whose roots satisfy $0 < y_i <y_{\rm max}$, with~$y_{N_y}$ sitting just below $y_{\rm max}$ and $y_{N_y+1}$ sitting just above.

\begin{figure}[t]
	\begin{center}
		\begin{subfigure}{.49\textwidth}
			\centering
			\includegraphics[width=\linewidth]{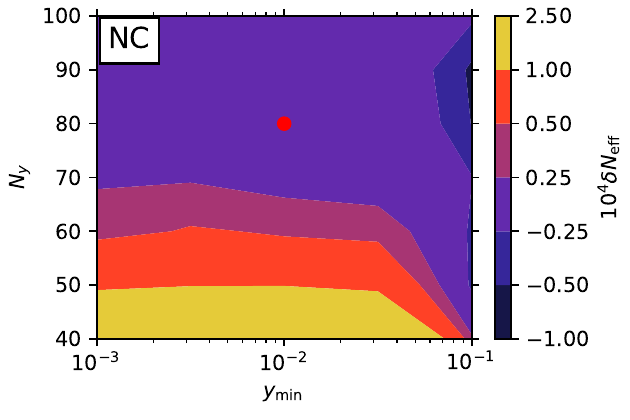}
		\end{subfigure}\\
		\begin{subfigure}{.49\textwidth}
			\centering
			\includegraphics[width=\linewidth]{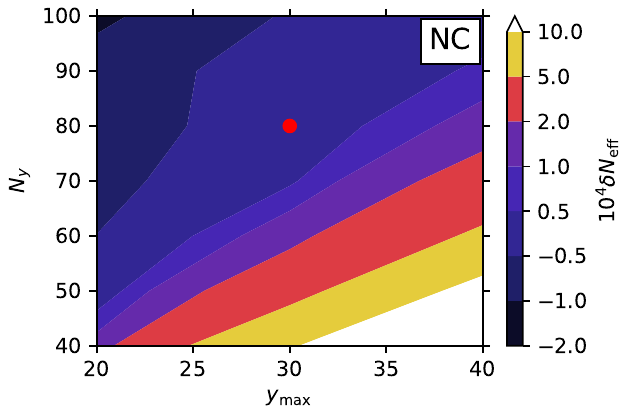}
		\end{subfigure}
		\begin{subfigure}{.49\textwidth}
			\centering
			\includegraphics[width=\linewidth]{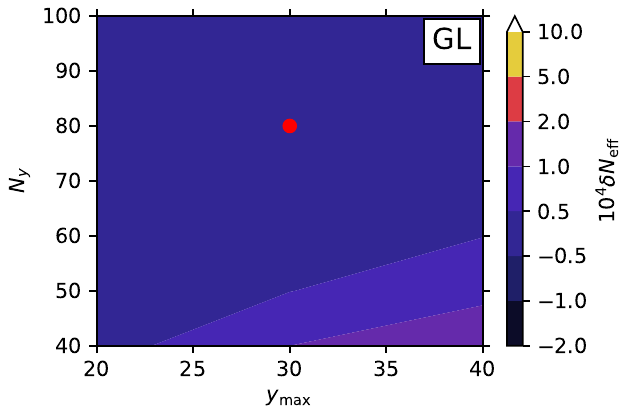}
		\end{subfigure}
	\end{center}
	\caption{Change in $\Neff^{\rm SM}$ in a full calculation including the complete neutrino--neutrino collision integral, with respect to variations in the number of $y$-nodes and their range.
	{\it Top}: Variations in $N_y$ and~$y_{\rm min}$ using the Newton--Cotes quadrature method.
	{\it Bottom left}: Variations in $N_y$ and~$y_{\rm max}$ using the Newton--Cotes quadrature method.
	{\it Bottom right}: Variations in $N_y$ and~$y_{\rm max}$ using the Gauss--Laguerre quadrature method.
	The setting used in our benchmark full calculation including the complete neutrino--neutrino collision integral ($y_{\rm min}=0.01$ where applicable, $y_{\rm max}=30$, and $N_y = 80$) is marked by a red dot on each plot.
	\label{fig:numerical}}
\end{figure}

This particular method of $y_i$ selection also means that the actual value of the highest~$y$ always fluctuates below the nominal
$y_{\rm max}$, at a distance depending on both our choices of $y_{\rm max}$ and $N_y$. Using the default setting of $y_{\rm max}=20$, table~\ref{tab:converge1} shows the outcome $\Neff$ under different set-ups {\it vis \`a vis} the neutrino--neutrino collision integral 
with respect to variations of $N_y$ in the range $N_y \in [40,80]$.
The bottom right panel of figure~\ref{fig:numerical} shows in addition variations with respect to our choice of $y_{\rm max}$ in the range $y_{\rm max} \in [20,40]$ for the full calculation.
Clearly, 
provided we do not decrease significantly the ratio $y_{\rm max}/N_y$, variations in $N_y$ and $y_{\rm max}$ within even fairly large ranges
do not generate differences larger than $|\delta \Neff| \sim 10^{-4}$, indicating that numerical convergence can be achieved at the $|\delta \Neff| \sim 10^{-4}$ level or better using GL quadrature.

\paragraph{Newton--Cotes (NC) quadrature} The NC quadrature method can be applied to any momentum grid, and we use for our benchmark $\Neff^{\rm SM}$ calculation 60 linearly-spaced nodes between $y_{\rm min}=0.01$ and $y_{\rm max}=20$ under the minimum set-up, and 
80 linearly-spaced nodes between $y_{\rm min}=0.01$ and $y_{\rm max}=30$ in a full calculation including the complete neutrino--neutrino collision.
As with the GL method, table~\ref{tab:converge1} and the top and bottom left panels of figure~\ref{fig:numerical} demonstrate that NC also performs at $|\delta \Neff| \sim {\cal O} (10^{-4})$ level or better with respect to variations in $N_y$ and $y_{\rm max}$, again provided we do not decrease $y_{\rm max}/N_y$ substantially.
In terms of variations in $y_{\rm min}$, figure~\ref{fig:numerical} 
shows our $\Neff$ estimates to be remarkably stable, shifting no more than $|\delta \Neff| \sim 2 \times 10^{-5}$ for a wide range of $y_{\rm min}$ values.%
\footnote{Note that these conclusions do not apply to $y_{\rm min}$ above 0.1 or $y_{\rm max}$ below 20, because the momentum range $0.1\lesssim y \lesssim20$ is the minimum that we need to sample of the neutrino momentum distribution, in order to capture 99.99\% of the (equilibrium) neutrino number density and 99.9995\% of the energy density.}

Note that figure~\ref{fig:numerical} is also useful for the purpose of finding the most economical momentum-grid configuration in order to achieve a required precision.
The execution time of \fortepiano\ does not scale linearly with $N_y$:
while the number of $y$-dependent differential equations that need to be solved does grow with $N_y$, 
the 2D integrals for the collision terms are computed on a $N_y\times N_y$ grid. Thus, 
being able to reduce the required $N_y$ to achieve a target precision can translate into a significant reduction in computation time.


\subsubsection{Initialisation time}

The neutrino density matrix $\varrho(x,y)$ is initialised at $x_{\rm in}$ with vanishing off-diagonal entries and diagonal entries equal to the ideal gas equilibrium occupation numbers at the rescaled temperature $z_{\rm in}$,
where $z_{\rm in}$ is found by evolving the continuity equation~\eqref{comoving_conserv_equation} from an even earlier time, $x_0 \equiv (1/10) (m_e/m_\mu)$, with $z_0 \equiv a_0 \times 10 \, m_\mu=1$, assuming equilibrium in the neutrino sector.
Our default choice of $x_{\rm in} = 0.001$ corresponds to $z_{\rm in} -1= 2.9 \times 10^{-4}$ or $T_{\rm in}\simeq 511$~MeV, a temperature at which the neutrino collision rate significantly dominates over both the flavour oscillation and the Hubble expansion rate (see, e.g., figure I of~\cite{Lunardini:2000fy}), so that the settings for $\varrho(x_{\rm in},y)$ hold to an excellent approximation.

\begin{figure}[t]
\begin{center}
\includegraphics[width=10cm]{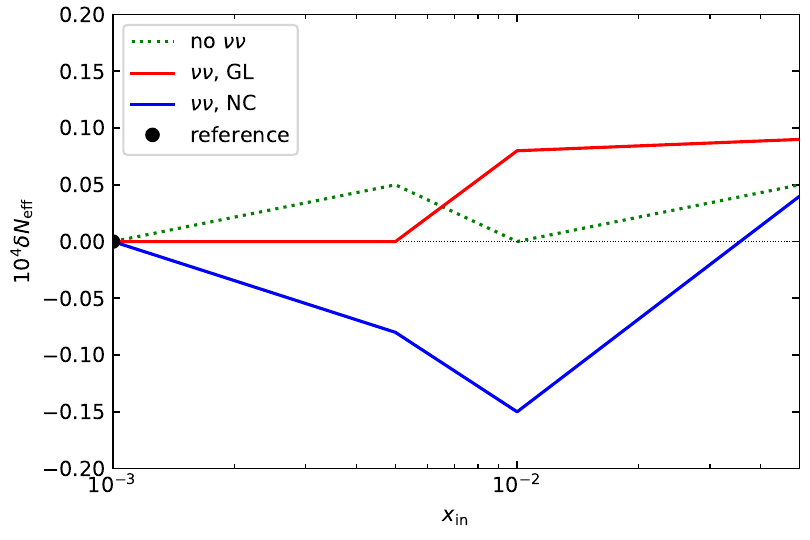}
\end{center}
\caption{Change in $\Neff^{\rm SM}$ with respect to variations in the initialisation time $x_{\rm in}$. The default choice is $x_{\rm in} = 0.001$. The solid red and blue lines denote the full calculation including the full neutrino--neutrino collision term using respectively the GL and the NC methods, while the green dotted line represents the minimum set-up computed with NC.\label{fig:xin}}
\end{figure}

To establish convergence with respect to this initialisation procedure, figure~\ref{fig:xin} shows the changes in $\Neff^{\rm SM}$ within the minimum set-up (dotted green) computed with NC and the full calculation including full neutrino--neutrino collisions using the GL (solid red) and the NC (solid blue) methods, as we systematically lower $x_{\rm in}$ over a range $x_{\rm in} \in [0.001, 0.05]$. The latter number corresponds to $z_{\rm in} = 1.098$%
\footnote{
This number comes from our choice of normalisation of $z$ and the fact that
we take into account the presence of muons at high temperatures.
When muons become non-relativistic, they transfer their entropy to the rest of the plasma,
with the result that the photon temperature grows by a factor
$(57/43)^{1/3}\simeq 1.098$.
}
or, equivalently, $T_{\rm in} \simeq 10$~MeV, technically still considerably above the nominal neutrino decoupling temperature, $T_d \sim {\cal O} (1)$~MeV.
Evidently, for the entire range of $x_{\rm in}$ tested, deviations from the default calculation are always below $|\delta \Neff| \sim 2 \times 10^{-5}$ and hence consistent with numerical noise. We therefore conclude that in the context of computing $\Neff^{\rm SM}$, the choice of initialisation time~$x_{\rm in}$ even as late as $x_{\rm in}=0.05$ is not a limiting factor.
This reaffirms the findings of~\cite{deSalas:2016ztq,Gariazzo:2019gyi}, which recommended $x_{\rm in} = 0.05$ for three-flavour decoupling calculations.


\subsection{Energy density and number density conservation tests}
\label{sec:energynumber}

As a final test of the convergence properties of \fortepiano,
we devise several controlled tests to quantify the extent to which \fortepiano\ conserves neutrino number and energy
under variations in the momentum discretisation described in section~\ref{sec:momentum}. Specifically, we solve the Boltzmann--continuity system under the following conditions for the collision integrals:
\begin{enumerate}
\item {\bf No collisions}, i.e., ${\cal I}_{\nu e}= {\cal I}_{\nu \nu}=0$.	
\item {\bf Neutrino--neutrino collisions only}, i.e., we set ${\cal I}_{\nu e}=0$, leaving only ${\cal I}_{\nu \nu}$ operative. 	
\item {\bf Neutrino--electron elastic scattering only}, i.e., we set ${\cal I}_{\nu \nu}=0$, as well as ${\cal I}_{\nu e}^{\rm ann}=0$ in equation~\eqref{eq:collint}.
\item{\bf No $e^+e^-$-annihilation}, i.e., we set ${\cal I}_{\nu e}^{\rm ann}=0$ in equation~\eqref{eq:collint}, while keeping all other collision terms operative.
\item {\bf Full collisions}, i.e., we take into account all the contributions to ${\cal I}_{\nu e}$ and ${\cal I}_{\nu \nu}$.
\end{enumerate}
Theoretically, we expect the comoving neutrino number density to be conserved in only the first four cases, while only the first two scenarios conserve the comoving energy density as well. Numerical detailed balance can be deemed well satisfied if the fractional number and, where appropriate, energy density losses in cases 1.-4.\ can be kept substantially below the target physical changes in the system represented by case 5.

\begin{figure}[t]
\begin{center}
\includegraphics[width=.7\textwidth]{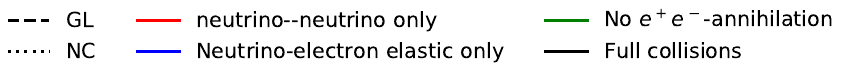}
\includegraphics[width=.49\textwidth]{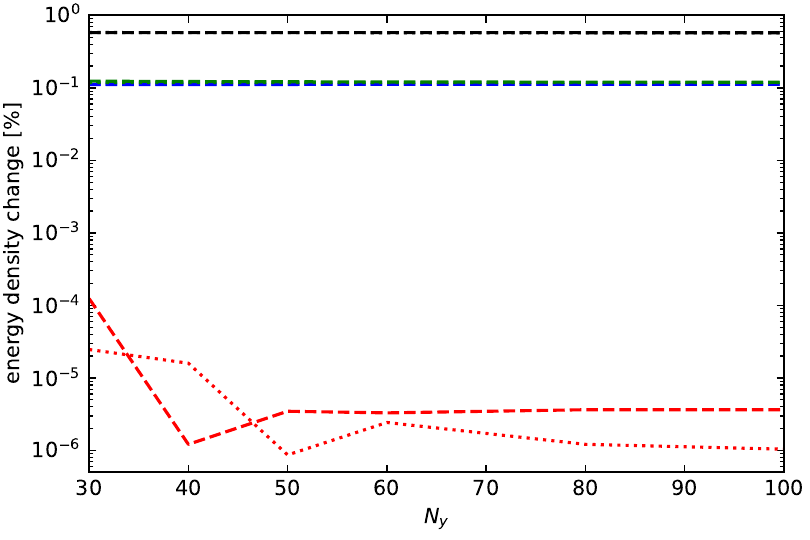}
\includegraphics[width=.49\textwidth]{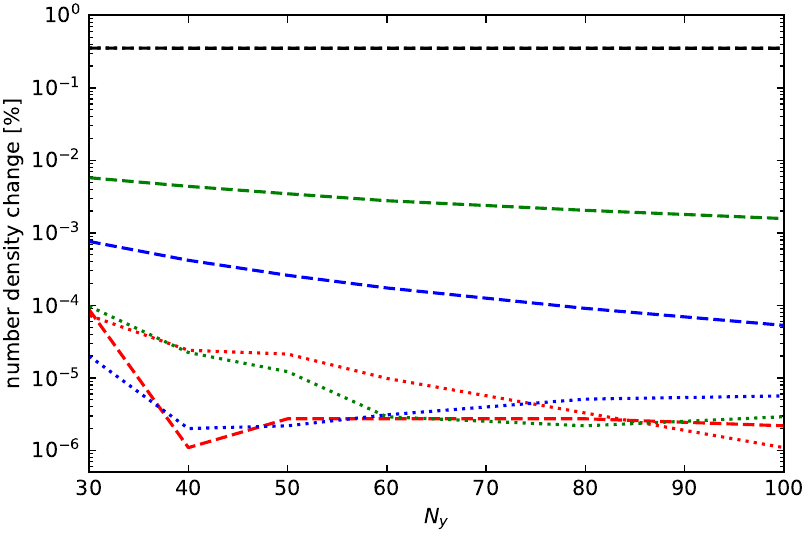}
\end{center}
\caption{{\it Left}:~Fractional change in percent in the comoving energy density relative to the initial state at $x_{\rm in}=0.001$, under variations of the quadrature method (Gauss--Laguerre vs Newton--Cotes), the number of $y$-nodes $N_y$, and modelling of the collision integral ${\cal I}[\varrho]$. {\it Right}:~Fractional change in percent in the comoving number density.	
		\label{fig:cons}}
\end{figure}

Figure~\ref{fig:cons} shows the fractional non-conservation in the comoving neutrino number and energy densities, defined respectively as the relative changes in the final-state comoving number and energy densities relative to the initial state. Evidently, where both number and energy conservation are expected (i.e., no collisions, and neutrino--neutrino collisions only), \fortepiano\ is able to replicate the outcome at $10^{-5}\%$ or better using both the Gauss--Laguerre and the Newton--Cotes quadrature schemes. Number conservation also holds at the same level for both methods when the energy-changing neutrino--electron collisions are switched on.

Interestingly, divergences between the Gauss--Laguerre and the Newton--Cotes scheme begin to appear when neutrino--electron interactions are turned on. 
Here, we see in figure~\ref{fig:cons} that while NC continues to incur non-conservation in the number density at the $<10^{-5}\%$ level, the amount of non-conservation is three orders of magnitude as large
under the GL scheme and does not converge under variation of $N_y$.
Nonetheless, at $\sim10^{-3}\%$ the degree of non-conservation of number density from numerical artefacts is still substantially below our accuracy goal.
We therefore do not deem the numerical breakdown of detailed balance to be of concern in our calculation, and estimate the numerical error incurred in the final $\Neff^{\rm SM}$ to be $|\delta \Neff| \sim 10^{-4}$, based on the convergence tests reported in table~\ref{tab:converge1} and figure~\ref{fig:numerical}.


\section{Physics results}\label{sec:results}

Having established that it is possible to achieve $|\delta \Neff| \sim 10^{-4}$ numerical convergence,
we are finally in a position to report our physics results. First and foremost, the new SM benchmark effective number of neutrinos has been determined to be
\begin{equation}
\label{eq:neffsm}
\Neff^{\rm SM} = 3.0440 \pm 0.0002
,
\end{equation}
under the conditions of
(i)~a normal neutrino mass ordering,
(ii)~finite-temperature QED corrections to the QED equation of state to ${\cal O}(e^3)$ and to the weak rates to ${\cal O}(e^2)$ of the type~(a) (see section~\ref{sec:weakrates} and figure~\ref{QED2to2_scatterings} for the classification),
and (iii) a full numerical treatment of the complete neutrino--electron and neutrino--neutrino collision integrals.
The quoted uncertainty is dominated by errors incurred in the numerical solution procedure ($|\delta \Neff|\sim10^{-4}$), followed by measurement uncertainties in the solar mixing angle $\sin^2\theta_{12}$ within its experimentally allowed $3\sigma$-range ($|\delta \Neff|\sim10^{-4}$).

\begin{table}[t]
	\centering
	\begin{tabular}{|l|c|c|c|}
		\hline
		& $\Neff^{\rm SM}$ (no osc)
		& $\Neff^{\rm SM}$ (NO)
		& $\Neff^{\rm SM}$ (IO)\\
		\hline
		\hline
		\multicolumn{2}{|c|}{{\bf Benchmark A} --- $\left\{{\cal I}_{\nu \nu}[\varrho]\right\}_{\alpha \alpha}=0$} \\
		\hline
		Assuming: 
		&	\parbox[t]{13mm}{\multirow{4}{*}{\rotatebox[origin=c]{0}{\bf 3.04263}}} 
		&	\parbox[t]{13mm}{\multirow{4}{*}{\rotatebox[origin=c]{0}{\bf 3.04360}}} 
		&	\parbox[t]{13mm}{\multirow{4}{*}{\rotatebox[origin=c]{0}{\bf 3.04361}}} \\
		\textbullet~$(2)\slashed{\ln}+(2)\ln+(3)+$ type (a) weak rates &&& \\
		\textbullet~Damping for $\left\{{\cal I}_{\nu e}[\varrho]\right\}_{\alpha \beta}$ &&& \\
		\textbullet~$N_y=60$, $y_{\rm max}=20$, NC linearly spaced $y_i$ &&& \\
		\hline
		\hline
		\multicolumn{4}{|c|}{\bf Alternative estimates} \\
		\hline
		\hline
		\multicolumn{4}{|c|}{Momentum grid} \\
		\hline
		$N_y=40$, $y_{\rm max}=20$, GL spacing of $y_i$ nodes
		&3.04261
		&3.04355
		&3.04360\\
		\hline
		\hline
		\multicolumn{4}{|c|}{Integrals for off-diagonal $\left\{{\cal I}_{\nu e}[\varrho]\right\}_{\alpha \beta}$} \\
		\hline
		$N_y=60$, $y_{\rm max}=20$, NC linearly spaced $y_i$
		&3.04261
		&3.04357
		&3.04362\\
		$N_y=40$, $y_{\rm max}=20$, GL spacing of $y_i$ 
		&3.04261
		&3.04357
		&3.04364\\
		\hline
		\hline
		\multicolumn{4}{|c|}{Finite-temperature QED corrections} \\
		\hline
		$(2)\slashed{\ln}$
		&3.04361
		&3.04458
		&
		\\
		$(2)\slashed{\ln}+(2)\ln$
		&3.04358
		&3.04452
		&
		\\
		$(2)\slashed{\ln}+(3)$
		&3.04264
		&3.04361
		&
		\\
		$(2)\slashed{\ln}+(2)\ln+(3)$
		&3.04263
		&3.04360
		&
		\\
		\hline
	\end{tabular}
	\caption{Summary of our $\Neff^{\rm SM}$ estimates within the minimum set-up (i.e., no diagonal neutrino--neutrino collision integral),	computed for different settings of neutrino flavour oscillations and finite-temperature QED corrections.
		The labels $(2) \ln$, $(2)\slashed{\ln}$, and $(3)$ denote respectively the ${\cal O}(e^2)$ logarithmic,
		${\cal O}(e^2)$ log-independent, and ${\cal O}(e^3)$ finite-temperature corrections to the QED equation of state.}
	\label{tab:Results_nonunudiag}
\end{table}

Table~\ref{tab:Results_nonunudiag} summarises the contributions of various elements to $\Neff^{\rm SM}$ computed within the minimum set-up, while table~\ref{tab:Results_nunudiag} reports those calculations that incorporate diagonal elements of the neutrino--neutrino collision integral, $\left\{{\cal I}_{\nu \nu}[\varrho]\right\}_{\alpha \alpha}$, implemented in different ways.
We shall discuss these results in more detail below. Note however in table~\ref{tab:Results_nonunudiag} that switching the neutrino mass ordering from normal to inverted generates but a negligible $|\delta \Neff| \sim 10^{-5}$ difference in~$\Neff^{\rm SM}$. Henceforth, we shall concern ourselves mainly with a normal mass ordering.



\subsection{New finite-temperature QED corrections}\label{res:QED}

Relative to the 2016 calculations of~\cite{deSalas:2016ztq}, table~\ref{tab:Results_nonunudiag} shows that the main modification to $\Neff^{\rm SM}$ riginates in
the $\mathcal{O}(e^3)$ finite-temperature correction to the QED equation of state. This correction generates a change of $\delta \Neff \sim -0.001$ under the Benchmark A settings,
bringing the 2016 value of 3.045~\cite{deSalas:2016ztq} down to 3.0440,
consistent with the estimate of~\cite{Bennett:2019ewm} within the instantaneous decoupling approximation.

We further test the impact of various combinations of finite-temperature QED corrections on $\Neff^{\rm SM}$, also shown in table~\ref{tab:Results_nonunudiag}.
These numbers are again largely consistent with the instantaneous-decoupling estimates of~\cite{Bennett:2019ewm}, and reaffirm the conclusions of~\cite{Bennett:2019ewm}
that the ${\cal O}(e^2)$ logarithmic correction to the QED equation of state ($\delta \Neff \sim - 4 \times 10^{-5}$) and the weak rate corrections of type~(a) ($|\delta \Neff| \lesssim 10^{-4}$) are strictly not necessary to achieve a prediction of $\Neff^{\rm SM}$ to four-digit significance.


\subsection{Full collision integral for neutrino--neutrino scattering}
\label{sec:fullnunu}

\begin{table}[t]
	\centering
	\begin{tabular}{|l|c|c|c|}
		\hline
		& $\Neff^{\rm SM}$ (no osc)
		& $\Neff^{\rm SM}$ (NO)
		& $\Neff^{\rm SM}$ (IO) \\
		\hline
		\hline
		\multicolumn{4}{|c|}{{\bf Benchmark B} --- $\left\{{\cal I}_{\nu \nu}[\varrho]\right\}_{\alpha \alpha} \neq 0$} \\
		\hline
		Assuming: &	\parbox[t]{12mm}{\multirow{4}{*}{\rotatebox[origin=c]{0}{\bf 3.04341}}}
		& \parbox[t]{12mm}{\multirow{4}{*}{\rotatebox[origin=c]{0}{\bf 3.04398}}} 
		& \parbox[t]{12mm}{\multirow{4}{*}{\rotatebox[origin=c]{0}{\bf 3.04399}}} \\
		\textbullet~$(2)\slashed{\ln}+(2)\ln+(3)+$ type (a) weak rates & & & \\
		\textbullet~Full ${\cal I}_{\nu e}[\varrho]$ and ${\cal I}_{\nu \nu}[\varrho]$ & & & \\
		\textbullet~$N_y=80$, $y_{\rm max}=30$, NC linearly spaced $y_i$ & & & \\
		\hline
		\hline
		\multicolumn{4}{|c|}{\bf Alternative estimates} \\
		\hline
		\hline
		\multicolumn{4}{|c|}{Momentum grid} \\
		\hline
		$N_y=80$, $y_{\rm max}=30$, GL spacing of $y_i$
		& 3.04334
		& 3.04392
		& 3.04392 \\
		$N_y=80$, $y_{\rm max}=20$, NC linearly spaced $y_i$
		& 3.04334
		& 3.04389
		& 3.04391 \\
		$N_y=80$, $y_{\rm max}=20$, GL spacing of $y_i$
		& 3.04334
		& 3.04386
		& 3.04393 \\
		\hline
		\hline
		\multicolumn{4}{|c|}{Off-diagonal collision terms} \\
		\hline
		Damping terms, NC quadrature
		& 
		& 3.04408
		& 
		\\
		Damping terms, GL quadrature
		& 
		& 3.04399
		& 
		\\
		\hline
		\hline
		\multicolumn{4}{|c|}{Neutrino--neutrino collision integral - $y_{\rm max}=20$} \\
		\hline
		Diagonal $\varrho$
		& 3.04333
		& 3.04416
		& 
		\\
		Full $\varrho$, interpolate $\varrho$/FD only in diagonal
		& 3.04334
		& 3.04389
		& 
		\\
		Full $\varrho$, interpolate $\varrho$/FD also in off-diagonal
		& 3.04334
		& 3.04389
		& 
		\\
		\hline
	\end{tabular}
	\caption{Summary of our $\Neff^{\rm SM}$ estimates, obtained for various implementations of the diagonal neutrino--neutrino collision integral.
		We report results with no neutrino oscillations (no osc), with oscillations assuming a normal mass ordering (NO) or an inverted mass ordering~(IO) and the central neutrino mixing parameter values of table~\ref{tab:uncertainties}.}
	\label{tab:Results_nunudiag}
\end{table}

Evaluating the full neutrino--neutrino collision integral in the computation of $\Neff^{\rm SM}$,
especially its diagonal entries $\{ {\cal I}_{\nu\nu}[\varrho]\}_{\alpha \alpha}$, raises the number by $\delta \Neff \simeq 4 \times 10^{-4}$ to $\Neff^{\rm SM} = 3.0440$. This is the Benchmark B value reported in table~\ref{tab:Results_nunudiag} including neutrino oscillations, and also our final recommended SM benchmark presented in equation~\eqref{eq:neffsm}.
The choice between using the full neutrino density matrix $\varrho(y)$ (``full'') or only its diagonal elements (``diagonal $\varrho$'') in the evaluation of the diagonal entries $\{ {\cal I}_{\nu\nu}[\varrho]\}_{\alpha \alpha}$ impacts on the final outcome at a comparable, $|\delta \Neff| \simeq 2 \times 10^{-4}$ level.%
\footnote{The diagonal $\varrho$ approximation combined with no flavour oscillations is equivalent to the scenario considered in references~\cite{Birrell:2014uka,Grohs2016,Froustey:2019owm}. The $\Neff$ values reported in these works are consistently larger than our no oscillations Benchmark B of $3.04341$ by about $0.001$, because they did not include the ${\cal O}(e^3)$ finite-temperature correction to the QED equation of state in their computation.}
We therefore recommend that a full evaluation of $\{ {\cal I}_{\nu\nu}[\varrho]\}_{\alpha \alpha}$ be employed whenever neutrino oscillations are switched on.

Another point of note in table~\ref{tab:Results_nunudiag} is that including nonzero $\{ {\cal I}_{\nu\nu}[\varrho]\}_{\alpha \alpha}$ terms in the calculation
diminishes the net contribution of neutrino oscillations to the final outcome $\Neff^{\rm SM}$: Within the minimum set-up (i.e., $\{ {\cal I}_{\nu\nu}[\varrho]\}_{\alpha \alpha}=0$), oscillations add $\delta \Neff \sim 0.001$ to the no-oscillation estimate. With $\{ {\cal I}_{\nu\nu}[\varrho]\}_{\alpha \alpha} \neq 0$ this contribution consistently reduces to $\delta \Neff \sim 5 \times 10^{-4}$. This is a physically sensible result: allowing neutrinos to scatter amongst themselves reduces the reliance of the system on oscillations as a means of energy transport between different flavours. From the perspective of computing
$\Neff^{\rm SM}$ to four-digit significance, this result indicates that
while flavour oscillations are not as crucial an ingredient as once envisaged, they are nonetheless of sufficient importance to include in a calculation to avoid rounding errors.


\subsection{Assessment of remaining uncertainties}

We have seen at the end of section~\ref{sec:energynumber} that the uncertainty incurred in $\Neff^{\rm SM}$ from our numerical solution procedure is estimated to be $|\delta \Neff| \sim 10^{-4}$.
The next finite-temperature correction to the QED equation of state is expected to enter at $\delta \Neff \sim 4 \times 10^{-6}$~\cite{Bennett:2019ewm}. Finite-temperature QED corrections to the weak rates of the types~(b),~(c) and~(d), on the other hand, have yet to be determined, although one might reasonably guess that their contributions to $\Neff^{\rm SM}$ are similar in magnitude to that of the type (a), which sits at just below $\delta \Neff \sim 10^{-4}$ according to table~\ref{tab:Results_nonunudiag}. Other sources of uncertainties investigated and their impact on $\Neff^{\rm SM}$ are presented in tables~\ref{tab:Results_nonunudiag} and \ref{tab:Results_nunudiag} as well as figure~\ref{fig:oscillation}, on which we elaborate below.


\subsubsection{Approximate treatment of the weak collision integrals}

Approximate treatments of the diagonal neutrino--neutrino collision integral and their impact on $\Neff^{\rm SM}$ as detailed in table~\ref{tab:Results_nunudiag} have already been discussed in section~\ref{sec:fullnunu}. To paraphrase the conclusion, setting $\{ {\cal I}_{\nu\nu}[\varrho]\}_{\alpha \alpha}=0$ generally underestimates $\Neff^{\rm SM}$ by about $\delta \Neff \simeq 4 \times 10^{-4}$,
while approximating the full density matrix $\varrho$ with only its diagonal entries in the evaluation of $\{ {\cal I}_{\nu\nu}[\varrho]\}_{\alpha \alpha}$ 
(the ``diagonal $\varrho$'' approximation) generates a comparable difference of $|\delta \Neff| \simeq 2 \times 10^{-4}$ and should hence be avoided when neutrino oscillations are switched on.

Of lesser impact is the exact modelling of
the {\it off-diagonal} entries of the neutrino--electron and neutrino--neutrino collision integrals, $\{{\cal I}_{\nu e}[\varrho]\}_{\alpha \beta}$ and $\{{\cal I}_{\nu \nu}[\varrho]\}_{\alpha \beta}$, when neutrino oscillations are operative (these terms are irrelevant in the no-oscillation case).
While this is one of the first works to use a full evaluation of
$\left\{{\cal I}_{\nu \nu}[\varrho]\right\}_{\alpha\beta}$ in the computation of~$\Neff^{\rm SM}$, we find that substituting $\left\{{\cal I}_{\nu \nu}[\varrho]\right\}_{\alpha\beta}$  and $\left\{{\cal I}_{\nu e}[\varrho]\right\}_{\alpha\beta}$ with the damping approximation~\eqref{eq:damping} generates a difference of $|\delta \Neff| \lesssim 10^{-4}$.
We therefore conclude that the off-diagonal damping approximation may be a reasonable compromise if computing resources need to be diverted elsewhere.


\subsubsection{Measurement errors in the physical parameters of the neutrino sector}
\label{ssec:weak}

\begin{figure}
\centering
\includegraphics[width=0.5\textwidth]{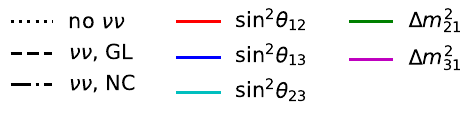}

\includegraphics[width=0.48\textwidth]{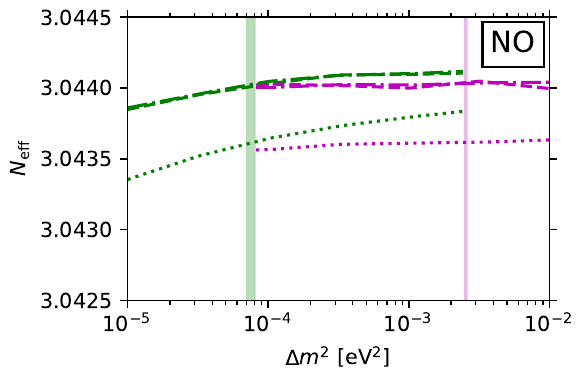}
\includegraphics[width=0.48\textwidth]{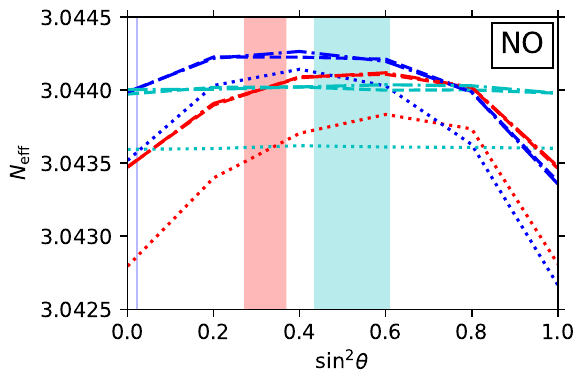}

\includegraphics[width=0.48\textwidth]{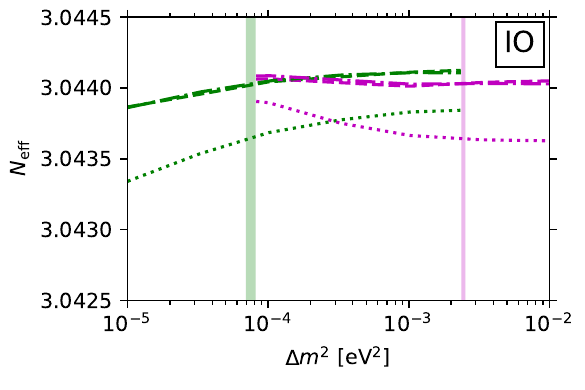}
\includegraphics[width=0.48\textwidth]{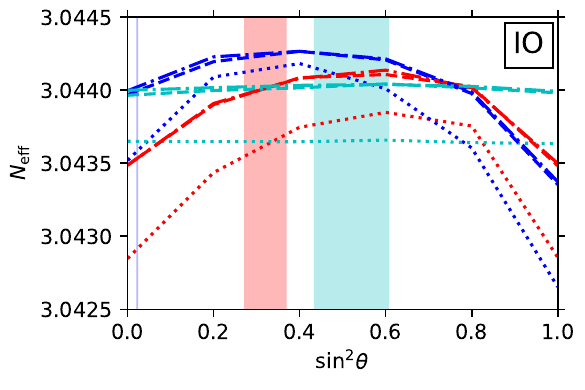}
\caption{Change in $\Neff^{\rm SM}$ under variations in the physical parameters of the neutrino sector, assuming a normal mass ordering (top row) and an inverted mass ordering (bottom row). We consider variations in the mass splittings $\Delta m^2_{21}$ and $\Delta m^2_{31}$, assuming at all times $\Delta m^2_{21}<|\Delta m^2_{31}|$ (left column), and in the mixing angles $\sin^2 \theta_{12}$, $\sin^2 \theta_{13}$, and $\sin^2 \theta_{23}$ (right column). We show results for both the minimum set-up (dotted lines) and the full calculation (dot-dash lines).
Coloured bands represent the current $3\sigma$-range of each parameter given in table~\ref{tab:uncertainties}~\cite{deSalas:2020pgw}.
\label{fig:oscillation}
}
\end{figure}

Figure~\ref{fig:oscillation} shows changes in the predicted $\Neff^{\rm SM}$ with respect to variations in the neutrino mass splittings and mixing parameters used in the calculation, for both a normal and an inverted neutrino mass ordering.
We consider these variations within the minimum set-up where $\{ {\cal I}_{\nu\nu}[\varrho]\}_{\alpha \alpha}=0$, as well as in the context of a full, $\{ {\cal I}_{\nu\nu}[\varrho]\}_{\alpha \alpha} \neq 0$ calculation.

First of all, we note that the minimum set-up is, in comparison with the full calculation, visibly more sensitive to variations in the neutrino parameters. Ramping up $\sin^2 \theta_{12}$ from zero to its best-fit value of $0.32$, for example, generates $\delta \Neff \simeq 0.001$ in the minimal set-up and $\delta \Neff \simeq 0.0005$ in the full calculation.
This result is consistent with our earlier observation in section~\ref{sec:fullnunu} that including flavour oscillations in the calculation has a stronger impact on the outcome $\Neff^{\rm SM}$ within the minimum set-up than within the full calculation.

The actual variations in $\Neff^{\rm SM}$ with respect to changes in the parameter values --- especially if we
restrict our attention to the $3\sigma$ regions (shaded regions) --- are however generally quite small.
Variations with respect to the neutrino mass splittings, notwithstanding the relatively large, percent-level $1\sigma$-uncertainties in $\Delta m^2_{21}$ and $\Delta m^2_{31}$, are practically indiscernible in the $3\sigma$ region.
Physically, such insensitivity of $\Neff^{\rm SM}$ to $\Delta m^2_{21}$ and $\Delta m^2_{31}$ reflects the fact that the typical oscillation frequency, $\Delta m^2_{ij}/2 \langle p \rangle$, in all oscillation channels is much larger than the Hubble expansion rate across the neutrino decoupling epoch (see, e.g., figure I of~\cite{Lunardini:2000fy}).
This means that the oscillation probability averaged over a Hubble time is effectively dependent only on the mixing angle; the exact mass splittings largely drop out of the picture.

For the neutrino mixing angles, we likewise find no discernible dependence of $\Neff^{\rm SM}$ on $\sin^2\theta_{23}$ across the whole parameter range tested ($\sin^2 \theta_{23} \in [0,1]$),
clearly because this oscillation channel merely swaps the $\nu_\mu$ and the $\nu_\tau$ populations. Except for a largely inconsequential
dependence on the muon energy density in the matter potential part of the flavour oscillations Hamiltonian~\eqref{eq:hamiltonian},%
\footnote{
The $\nu_{\mu} \leftrightarrow \nu_{\tau}$ vacuum oscillation frequency typically supersedes the Hubble expansion rate at $T \sim 20$~MeV, by which time the muon energy density has become strongly Boltzmann-suppressed.
For this reason, the muon energy density that appears in the matter potential is largely inconsequential for the computation of $\Neff$.
}
these two populations are, for the $\Neff^{\rm SM}$ calculation, essentially identical.
Changing the rate at which they are swapped therefore has no real impact on the outcome.

The only instances in which we find a reasonably strong, $\delta \Neff \sim {\cal O}(10^{-4})$ response of $\Neff^{\rm SM}$ to parameter variations are the cases of $\sin^2 \theta_{12}$ and $\sin^2 \theta_{13}$, where tuning these parameters towards
$\sin^2\theta_{ij} \simeq 0.5$ tends to increase $\Neff^{\rm SM}$.
These mixing angles control the oscillations between the $\nu_e$ and the $\nu_{\mu,\tau}$ populations, the former of which has the important feature that it decouples from the QED plasma last because of the charged-current $\nu_e e$-interactions unique to it.
Then, a larger mixing angle enhances the energy transfer rate between $\nu_e$ and $\nu_{\mu,\tau}$, and
keeps the latter populations effectively coupled to the QED plasma for a longer time to partake in the entropy transfer from $e^+e^-$-annihilation.

For a $\pm 3 \sigma$ ($\sim 20$\%) variation in $\sin^2 \theta_{12}$ from its central value, figure~\ref{fig:oscillation} shows that $\Neff^{\rm SM}$ can change by as much as $\delta \Neff \sim \pm 10^{-4}$, consistent with expectations (i.e., $\sim 20$\% of the change due to flavour oscillations $\delta \Neff \sim 5 \times 10^{-4}$). For similar relative variations in $\sin^2 \theta_{13}$, the change in $\Neff^{\rm SM}$ is an order of magnitude smaller. But this is merely because the central value of $\sin^2 \theta_{13}$ itself is an order of magnitude smaller than $\sin^2 \theta_{12}$ to begin with, and hence plays only a subdominant role to the latter in terms of facilitating energy transfer between the $\nu_e$ and the $\nu_{\mu,\tau}$ populations. Had it been possible to increase $\sin^2 \theta_{13}$ to $\sim 0.5$, figure~\ref{fig:oscillation} shows that $\Neff^{\rm SM}$ would have been enhanced by $\delta \Neff \sim 2 - 3 \times 10^{-4}$.


\section{Conclusions}\label{sec:conclusions}

We have updated in this work the standard-model benchmark value for the effective number of neutrinos, $\Neff^{\rm SM}$, that quantifies the cosmological neutrino-to-photon energy densities, and estimated its uncertainties. Our recommended value of $\Neff^{\rm SM} = 3.0440 \pm 0.0002
$
has been established through a careful tracking of relic neutrino decoupling in the presence of neutrino flavour oscillations (assuming a normal neutrino mass ordering), as well as finite-temperature effects in the QED plasma.
The error estimate takes into account numerical uncertainty at the level of $|\delta \Neff| \sim 10^{-4}$ due to the discretisation of the momentum grid, plus a physical error of  $|\delta \Neff|  \sim 10^{-4}$ arising from the current $3\sigma$ measurement uncertainty in the neutrino oscillation parameters.
Our central $\Neff^{\rm SM}$ value is in perfect agreement with the calculation of reference~\cite{Froustey:2020mcq} which incorporated the same physics; our nominal theoretical uncertainty is somewhat larger than that quoted in~\cite{Froustey:2020mcq}, however, owing to different accounts of the physical error from the oscillation parameters.

Relative to the 2016 calculation~\cite{deSalas:2016ztq} which has a nominal uncertainty of $\pm 0.001$,
 the benchmark has shifted by $\delta \Neff \simeq -0.001$ because of a hitherto neglected ${\cal O}(e^3)$ finite-temperature correction to the QED equation state.
 We note that the oft-quoted 2005 number $\Neff^{\rm SM} = 3.046$~\cite{Mangano:2005cc} also misses the ${\cal O}(e^3)$ correction and has a nominal uncertainty of $\pm 0.002$~\cite{Mangano:2005cc}. A leading-digit breakdown of the various SM effects that contribute to $\Neff^{\rm SM}$'s deviation from 3 is presented in table~\ref{tab:Split}.

\begin{table}[t]
\centering
\begin{tabular}{|l|c|}
\hline
Standard-model corrections to $\Neff^{\rm SM}$ & Leading-digit contribution \\
\hline
$m_e/T_d$ correction& $+0.04$ \\
$\mathcal{O}(e^2)$ FTQED correction to the QED EoS& $+0.01$\\
Non-instantaneous decoupling+spectral distortion & $-0.005$\\
$\mathcal{O}(e^3)$ FTQED correction to the QED EoS& $-0.001$\\
Flavour oscillations & $+0.0005$\\
Type (a) FTQED corrections to the weak rates & $\lesssim 10^{-4}$\\
\hline
\end{tabular}
\caption{Leading-digit contributions from various SM corrections, in order of importance, accounted for in this work that make up the final $\Neff^{\rm SM}-3$. The largest, $m_e/T_d$ correction is identically the $\slashed{\rm Rel}$ correction discussed in~\cite{Bennett:2019ewm}, while the non-instantaneous decoupling+spectral distortion correction is defined relative to an estimate of $\Neff^{\rm SM}$ in the limit of instantaneous decoupling assuming $T_d = 1.3453$~MeV~\cite{Bennett:2019ewm}.}
\label{tab:Split}
\end{table}

The uncertainty in the benchmark $\Neff^{\rm SM}$ we quote in equation~\eqref{eq:neffsm} is a conservative sum of the two broad classes of errors examined in this work: numerical convergence of the solution procedure, notably momentum discretisation, and measurement uncertainties in the physical parameters of the neutrino sector.
It is chiefly dominated by the former, and is further augmented by contributions from measurement errors in the solar neutrino mixing angle $\sin^2 \theta_{12}$. Other uncertainties, including higher-order finite-temperature QED corrections, measurement errors in the other oscillation parameters, transients, etc., all fall below the intrinsic numerical noise of \fortepiano, which, in the context of computing $\Neff^{\rm SM}$, is in the ballpark of $|\delta \Neff| \sim 10^{-5}$.

Relative to the nominal uncertainty of $\delta \Neff \sim \pm 2 \times 10^{-4}$ given in equation~\eqref{eq:neffsm}, we believe we have exhausted all possible effects within the standard model of particle physics
that would change~$\Neff^{\rm SM}$ appreciably. An uncertainty of this magnitude also more than suffices to minimise the total error budget in the inference of cosmological parameters from the forthcoming generation of cosmological observations~\cite{Abazajian:2016yjj}.
Nevertheless, even if only for completeness, estimates of the types~(b),~(c) and~(d) finite-temperature QED corrections to the weak rates remain on the table, and there is certainly scope for beating down the uncertainty in~$\Neff^{\rm SM}$ even further.
In the latter regard,
the measurement of $\sin^2 \theta_{12}$, for example, is expected to be improved by the next generation of neutrino oscillations experiment
to a better-than-1\% determination~\cite{An:2015jdp} (see also \cite{Ellis:2020hus}).
A dedicated investigation of the stability of the weak collision integral evaluation could potentially also eliminate a large and currently dominating chunk in the nominal uncertainty in~$\Neff^{\rm SM}$.
We leave these for future work.

\acknowledgments

JJB and Y$^3$W are supported in part by the Australian Government through the Australian Research Council's Discovery Project (project DP170102382) and Future Fellowship (project FT180100031) funding schemes.
GB acknowledges the support of the National Fund for Scientific Research (F.R.S.- FNRS Belgium) through a FRIA grant.
MaD and Y$^3$W acknowledge support from the ASEM-DUO fellowship programme of the Belgian Acad\'emie de recherche et d'enseignement sup\'erieur (ARES).
PFdS acknowledges support by the Vetenskapsr{\aa}det (Swedish Research Council) through contract No. 638-2013-8993 and the Oskar Klein Centre for Cosmoparticle Physics.
SG and SP are supported by the Spanish grants FPA2017-85216-P (AEI/FEDER, UE), PROMETEO/2018/165 (Generalitat Valenciana) and the Red Consolider MultiDark FPA2017-90566-REDC.
SG acknowledges financial support by the ``Juan de la Cierva-Incorporaci\'on'' program (IJC2018-036458-I) of the Spanish MICINN until September 2020, from the European Union's Horizon 2020 research and innovation programme under the Marie Skłodowska-Curie grant agreement No 754496 (project FELLINI) starting from October 2020, and thanks the Institute for Nuclear Theory at the University of Washington for its hospitality and the Department of Energy for partial support during the preparation of this work.
We thank Julien Froustey and Oleksandr Tomalak for pointing out an error in the original manuscript.


\appendix
\section{Collision integrals}
\label{sec:collisionintegralsapp}

Accounting only for $2 \to 2$ neutrino--electron and neutrino--neutrino collision processes, the weak collision integral~${\cal I}[\varrho(x,y)]$ splits naturally into two parts, ${\cal I}[\varrho(x,y)]= {\cal I}_{\nu e}[\varrho(x,y)]+ {\cal I}_{\nu \nu}[\varrho(y)]$. At tree level, their general forms can be found in, e.g.,~\cite{Sigl:1992fn}. Here, we give their 2D-reduced forms under the assumptions of (i)~spatial homogeneity and isotropy, and (ii)~$CP$-symmetry.
The integral reduction follows the procedure of~\cite{Dolgov:1997mb}. We note however other reduction methods exist~\cite{Hannestad:2015tea,Bennett:2019ewm}, which yield formally different but numerically identical results.

\paragraph{Neutrino--electron.} The neutrino--electron collision integral ${\cal I}_{\nu e}[\varrho(x,y)]$ splits further into a scattering and an annihilation part,
\begin{equation}
\mathcal{I}_{\nu e}[\varrho(x,y)]
=
\frac{G_F^2}{(2\pi)^3y^2}
\left\{I^{\rm sc}_{\nu e}[\varrho(x,y)] + I^{\rm ann}_{\nu e}[\varrho(x,y)]\right\}\,,
\label{eq:collint}
\end{equation}
with
\begin{equation}
\begin{aligned}
I^{\rm sc} _{\nu e}
&=
\int {\rm d}y_2 {\rm d}y_3 \frac{y_2}{\epsilon_2}
\label{eq:I_sc}
\\
&
\times \left\{\left(\Pi_{2a}^{\rm sc}(y, y_2)+\Pi_{2b}^{\rm sc}(y, y_4)\right)
\left[
F_{\rm sc}^{LL}\left(\varrho^{(1)}, f_e^{(2)}, \varrho^{(3)}, f_e^{(4)}\right)
+F_{\rm sc}^{RR}\left(\varrho^{(1)}, f_e^{(2)}, \varrho^{(3)}, f_e^{(4)}\right)\right]\right.
\\
& \qquad \left.-2x^2\Pi_1^{\rm sc}(y,y_3)
\left[
F_{\rm sc}^{RL}\left(\varrho^{(1)}, f_e^{(2)}, \varrho^{(3)}, f_e^{(4)}\right)
+F_{\rm sc}^{LR}\left(\varrho^{(1)}, f_e^{(2)}, \varrho^{(3)}, f_e^{(4)}\right)
\right]
\right\}\,,
\end{aligned}\end{equation}
\begin{equation}
\begin{aligned}
I^{\rm ann}_{\nu e}
&=
\int {\rm d}y_2 {\rm d}y_3 \frac{y_3}{\epsilon_3}
\label{eq:l_ann}\\
&\times \left\{\Pi_{2b}^{\rm ann}(y, y_4)F_{\rm ann}^{LL}\left(\varrho^{(1)}, \varrho^{(2)}, f_e^{(3)}, f_e^{(4)}\right)
+\Pi_{2a}^{\rm ann}(y, y_3)F_{\rm ann}^{RR}\left(\varrho^{(1)}, \varrho^{(2)}, f_e^{(3)}, f_e^{(4)}\right)\right.
\\
&\qquad \left.+ x^2 \Pi_1^{\rm ann}(y,y_2)
\left[
F_{\rm ann}^{RL}\left(\varrho^{(1)}, \varrho^{(2)}, f_e^{(3)}, f_e^{(4)}\right)
+F_{\rm ann}^{LR}\left(\varrho^{(1)}, \varrho^{(2)}, f_e^{(3)}, f_{e}^{(4)}\right)
\right]
\right\}\,,
\end{aligned}
\end{equation}
and $\epsilon^2_i = x^2+y_i^2$ is a rescaled energy.
The scattering kernels are given by
\begin{equation}
\begin{aligned}
\Pi_1^{\rm sc}(y,y_3)
&=
y\,y_3\,D_1+D_2(y,y_3,y_2,y_4),
\\
\Pi_1^{\rm ann}(y,y_2)
&=
y\,y_2\,D_1-D_2(y,y_2,y_3,y_4),
\\
\Pi_{2a}^{\rm sc}(y,y_2)/2
&=
y\,\epsilon_2\,y_3\,\epsilon_4\,D_1 + D_3 - y\,\epsilon_2 D_2(y_3,y_4,y,y_2) - y_3\,\epsilon_4 D_2(y,y_2,y_3,y_4),
\\
\Pi_{2b}^{\rm sc}(y,y_4)/2
&=
y\,\epsilon_2\,y_3\,\epsilon_4\,D_1 + D_3 + \epsilon_2\,y_3 D_2(y,y_4,y_2,y_3) + y\,\epsilon_4 D_2(y_2,y_3,y,y_4),
\\
\Pi_{2a}^{\rm ann}(y,y_3)/2
&=
y\,y_2\,\epsilon_3\,\epsilon_4\,D_1 + D_3 + y\,\epsilon_3 D_2(y_2,y_4,y,y_3) + y_2\,\epsilon_4 D_2(y,y_3,y_2,y_4),
\\
\Pi_{2b}^{\rm ann}(y,y_4)/2
&=
y\,y_2\,\epsilon_3\,\epsilon_4\,D_1 + D_3 + y_2\,\epsilon_3 D_2(y,y_4,y_2,y_3) + y\,\epsilon_4 D_2(y_2,y_3,y,y_4),
\label{eq:scatteringkernels}
\end{aligned}
\end{equation}
where the functions $D_i$ are defined as follows~\cite{Dolgov:1997mb}:
\begin{equation}
\begin{aligned}
D_1(a,b,c,d)
&=
\frac{16}{\pi}
\int_0^\infty
\frac{{\rm d}\lambda}{\lambda^2}
\prod_{i=a,b,c,d}\sin(\lambda i)
\,,\\
D_2(a,b,c,d)
&=
-\frac{16}{\pi}
\int_0^\infty
\frac{{\rm d}\lambda}{\lambda^4}
\prod_{i=a,b}\left[\lambda i \cos(\lambda i)-\sin(\lambda i)\right]
\prod_{j=c,d}\sin(\lambda j)
\,,\\
D_3(a,b,c,d)
&=
\frac{16}{\pi}
\int_0^\infty
\frac{\mathrm{d}\lambda}{\lambda^6}
\prod_{i=a,b,c,d}\left[\lambda i \cos(\lambda i)-\sin(\lambda i)\right]
\,.
\end{aligned}
\end{equation}
All three integrals can be evaluated analytically. See, e.g.,~\cite{Blaschke:2016xxt} for the complete expressions.

Lastly, the phase space matrices $F_{\rm sc}^{ab}$ and $F_{\rm ann}^{ab}$, where $a,b= R,L$, are given by
\begin{equation}
\begin{aligned}
F_{\rm sc}^{ab}
\equiv&\,
f_e^{(4)}(\mathbb{1}-f_e^{(2)})G^a\varrho^{(3)}G^b(\mathbb{1}-\varrho^{(1)})
-f_e^{(2)}(\mathbb{1}-f_e^{(4)})\varrho^{(1)}G^b(\mathbb{1}-\varrho^{(3)})G^a + {\rm h.c.}\\
F_{\rm ann}^{ab}
\equiv&\,
f_e^{(3)}f_e^{(4)} G^a(\mathbb{1}-\varrho^{(2)})G^b(\mathbb{1}-\varrho^{(1)})
- (\mathbb{1}-f_e^{(3)})(\mathbb{1}-f_e^{(4)})G^a\varrho^{(2)}G^b\varrho^{(1)} + {\rm h.c.},
\label{eq:F_ab_ann}
\end{aligned}
\end{equation}
with $\varrho^{(i)}=\varrho(y_i)$, $f_e^{(i)}=f_e(y_i) \times\mathbb{1}$, and
\begin{equation}
\begin{aligned}
G^L=\text{diag}(g_L, \tilde g_L, \tilde g_L), \\
G^R=\text{diag}(g_R, g_R, g_R)\,
\end{aligned}
\end{equation}
are the weak coupling matrices for left- and right-handed particles, conventionally taken to be $g_L=\sin^2\theta_W+1/2$, $\tilde g_L=\sin^2\theta_W - 1/2$, and $g_R=\sin^2\theta_W$. In our calculation of $\Neff^{\rm SM}$, we use the values given in table~\ref{tab:uncertainties}
at zero-momentum transfer~\cite{Kumar:2013yoa,Erler:2013xha}.

\paragraph{Neutrino--neutrino.} The neutrino--neutrino collision integral likewise splits into a scattering and an annihilation part,
\begin{equation}
\mathcal{I}_{\nu \nu}[\varrho(x,y)]
=
\frac{G_F^2}{(2\pi)^3y^2}
\left\{I^{\rm sc}_{\nu \nu}[\varrho(x,y)] + I^{\rm pair}_{\nu \nu}[\varrho(x,y)]\right\}\,,
\label{eq:collintnunu}
\end{equation}
with
\begin{eqnarray}
I^{\rm sc} _{\nu \nu}
&=& \frac{1}{4}
\int {\rm d}y_2 {\rm d}y_3 \, \Pi_{2a}^{\rm sc}(y, y_2) \,
F_{\rm sc}^{\nu\nu}\left(\varrho^{(1)}, \varrho^{(2)}, \varrho^{(3)}, \varrho^{(4)}\right), \label{eq:Inunu_sc}\\
I^{\rm pair}_{\nu \nu} &=& \frac{1}{4}
\int {\rm d}y_2 {\rm d}y_3 \, \Pi_{2b}^{\rm sc}(y, y_4)\, F_{\rm pair}^{\nu\nu} \left(\varrho^{(1)}, \varrho^{(2)}, \varrho^{(3)}, \varrho^{(4)}\right).
\label{eq:Inunu}
\end{eqnarray}
The scattering kernels are identically those given above in equation~\eqref{eq:scatteringkernels}, and the phase space matrices are
\begin{equation}
\begin{aligned}
F_{\rm sc}^{\nu\nu} \equiv & \, (\mathbb{1}-\varrho^{(1)}) \varrho^{(3)} \! \left [(\mathbb{1}-\varrho^{(2)}) \varrho^{(4)} + {\rm tr}(\cdots) \right]
\!-\! \varrho^{(1)} (\mathbb{1}- \varrho^{(3)})\! \left [\varrho^{(2)} (\mathbb{1}- \varrho^{(4)} )+ {\rm tr}(\cdots) \right] + {\rm h.c.} ,\\
F^{\nu\nu}_{\rm pair}
\equiv& \, (\mathbb{1}-\varrho^{(1)}) (\mathbb{1}-
{\varrho}^{(2)}) \! \left [
{\varrho}^{(4)} \varrho^{(3)} + {\rm tr}(\cdots) \right]
\! -\! \varrho^{(1)}
{\varrho}^{(2)} \left [(\mathbb{1}-
{\varrho}^{(4)} )(\mathbb{1}- \varrho^{(3)} )+ {\rm tr}(\cdots) \right] \\
+&
(\mathbb{1}-\varrho^{(1)}) \varrho^{(3)} \! \left [
{\varrho}^{(4)}(\mathbb{1}-
{\varrho}^{(2)}) + {\rm tr}(\cdots) \right]
\! -\! \varrho^{(1)} (\mathbb{1}- \varrho^{(3)}) \! \left [(\mathbb{1}-
{\varrho}^{(4)} )
{\varrho}^{(2)} + {\rm tr}(\cdots) \right] + {\rm h.c.},
\label{eq:phasespace}
\end{aligned}
\end{equation}
where the notation ${\rm tr}(\cdots)$ denotes the trace of the preceding term.


\section{Damping approximation}
\label{sec:dampingcoefficients}

Observe that all collision integrals~\eqref{eq:I_sc}, \eqref{eq:l_ann}, \eqref{eq:Inunu_sc}, and~\eqref{eq:Inunu} come in the form
\begin{equation}
{\cal I}[\varrho(y)]= (\mathbb{1}- \varrho(y)) \Gamma_y^< - \varrho(y) \Gamma_y^> + {\rm h.c.},
\label{eq:gainloss}
\end{equation}
where $\Gamma_y^<$ and $\Gamma_y^>$ are the gain and loss terms respectively, and it is understood that they are $3 \times 3$ complex matrices.
Consider first an off-diagonal element, $\{{\cal I}[\varrho(y)]\}_{\alpha \beta}$, with $\alpha \neq \beta$. Defining $\Gamma_y \equiv \Gamma_y^< +\Gamma_y^>$ and writing out equation~\eqref{eq:gainloss} explicitly in index notation, we find
\begin{equation}
\big\{{\cal I}[\varrho(y)] \big\}_{\alpha \beta} =
- \big\{\varrho(y) \big\}_{\alpha \beta} \left[\big\{\Gamma_y \big\}_{\beta \beta}+\big\{\Gamma^*_y \big\}_{\alpha \alpha} \right]
- \sum_{\gamma \neq \beta} \big\{ \varrho(y) \big\}_{\alpha \gamma} \big\{\Gamma_y\big\}_{\gamma \beta}
-\sum_{\gamma \neq \alpha} \big\{ \varrho(y) \big\}_{\gamma \beta} \big\{\Gamma^*_y \big\}_{\gamma \alpha},
\label{eq:offdiagonal2}
\end{equation}
where we have used the fact that $\{\varrho(y) \}_{\alpha \beta} = \{\varrho(y)^* \}_{\beta \alpha}$, and summation over~$\gamma$ is implied. Observe that the two sums contain only off-diagonal entries of $\Gamma_y$.

Suppose now $\Gamma_y$ is diagonal. This might be the case if, for example, the density matrices $\varrho(y_2),\varrho(y_3),\varrho(y_4)$ that constitute the integrand of $\Gamma_y$ are all diagonal, or if their off-diagonal entries all oscillate with different phases so that $\{\Gamma_y\}_{\alpha \beta}$ integrates to zero. Then, equation~\eqref{eq:offdiagonal2} immediately simplifies to
\begin{equation}
\big\{{\cal I}[\varrho(y)] \big\}_{\alpha \beta}
\simeq - \big\{D(y) \big\}_{\alpha \beta} \big\{\varrho(y) \big\}_{\alpha \beta},
\label{eq:offdiagonal3}
\end{equation}
which is of a damping form, with damping coefficient $\{D(y)\}_{\alpha \beta} \equiv \{\Gamma_y\}_{\beta \beta}+\{\Gamma^*_y\}_{\alpha \alpha}$. This is the origin of the ``off-diagonal damping approximation''~\eqref{eq:damping}.

The exercise can be repeated also for a diagonal entry of ${\cal I}[\varrho(y)]$. Writing out equation~\eqref{eq:gainloss} in index form, we find
\begin{equation}
\big\{{\cal I}[\varrho(y)] \big\}_{\alpha \alpha} =
2 \big\{\Gamma_y^<\big\}_{\alpha \alpha}
-2 \big\{\varrho(y) \big\}_{\alpha \alpha} \big\{\Gamma_y \big\}_{\alpha \alpha} - \sum_{\gamma \neq \alpha} \Big[ \big\{ \varrho (y) \big\}_{\alpha \gamma} \big\{\Gamma_y\}_{\gamma \alpha} + \big\{ \varrho(y)^* \big\}_{\alpha \gamma}\big\{\Gamma^*_y \big\}_{\gamma \alpha} \Big],
\label{eq:diagonal1}
\end{equation}
where we have used the fact that $\{\varrho(y)\}_{\alpha \alpha}$ and $\{\Gamma_y\}_{\alpha \alpha}$ are real. Assuming again that all off-diagonal entries of $\Gamma_y$ are negligible, equation~\eqref{eq:diagonal1} simplifies to
\begin{equation}
\big\{{\cal I}[\varrho(y)] \big\}_{\alpha \alpha} \simeq
2 \big\{\Gamma_y^< \big\}_{\alpha \alpha}
-2 \big\{\varrho(y) \big\}_{\alpha \alpha} \big\{\Gamma_y \big\}_{\alpha \alpha} .
\label{eq:diagonal2}
\end{equation}
Lastly, if all particle species --- besides the one at the mode~$y$ represented by $\varrho(y)$ --- are in a state of thermal equilibrium, then detailed balance requires that $\Gamma_y^> = e^{\epsilon_y/z} \Gamma_y^<$, with a rescaled energy~$\epsilon_y$ associated with the mode~$y$, and a rescaled temperature~$z$.
Consequently, equation~\eqref{eq:diagonal2} can be recast in the form
\begin{equation}
\big\{{\cal I}[\varrho(y)] \big\}_{\alpha \alpha}
\simeq
- \big\{R(y) \big\}_\alpha
\Big[\big\{\varrho(y) \big\}_{\alpha \alpha} - f_{\rm eq}(y) \Big],
\label{eq:repop}
\end{equation}
where $\{R(y)\}_\alpha \equiv 2 \{\Gamma_y\}_{\alpha \alpha}$ is often called a ``repopulation coefficient''~\cite{McKellar:1992ja,Bell:1998ds, Hannestad:1995rs}, and for $\epsilon_y=y$ (valid for nearly massless neutrinos), $f_{\rm eq}(y)$ is simply the relativistic Fermi--Dirac distribution. Equation~\eqref{eq:repop}, then, is the ``diagonal damping approximation''.

Comparing the definitions of the diagonal and off-diagonal damping coefficients, $\{R(y)\}_\alpha$ and $\{D(y)\}_{\alpha \beta}$, we identify the relation
\begin{equation}
\big\{D(y) \big\}_{\alpha \beta} = \frac{1}{2} \left[\big\{R(y) \big\}_\alpha+ \big\{R(y) \big\}_\beta \right],
\end{equation}
which is well known especially in the context of $\alpha$ and $\beta$ representing respectively an active and a sterile neutrino flavour.
Then, combining equations~\eqref{eq:offdiagonal3} and~\eqref{eq:diagonal2}, we can now write down a ``general damping approximation'',
\begin{equation}
\big\{{\cal I}[\varrho(y)]\big\}_{\alpha \beta} = - \big\{ D(y)\big\}_{\alpha \beta} \Big[\big\{\varrho(y) \big\}_{\alpha \beta} - \delta_{\alpha \beta}f_{\rm eq}(y) \Big],
\label{eq:general}
\end{equation}
for both diagonal $\alpha = \beta$ and off-diagonal $\alpha \neq \beta$ entries of a collision integral.

Before proceeding to derive the repopulation and damping coefficients, note that while for completeness we have given the general damping approximation~\eqref{eq:general}, in practical implementations the diagonal version typically does not have sufficient accuracy relative to our goals, owing to the difficulty in knowing the correct form of $f_{\rm eq}(y)$ to which the neutrino ensemble should tend. Therefore, the minimum set-up of our calculation always solves the diagonal entries of the neutrino--electron collision integral~${\cal I}_{\nu e}[\varrho(y)]$ in full. Note however that the approximate damping scheme proposed in reference~\cite{Hannestad:2015tea} may circumvent this problem.


\subsection{Damping coefficients}

\paragraph{Neutrino--neutrino.}
To evaluate the repopulation and hence damping coefficients corresponding to the neutrino--neutrino collision integral~${\cal I}_{\nu \nu}[\varrho(y)]$, we first assume that $\varrho(y') = f_{\rm eq}(y') \mathbb{1}$ for all $y'\neq y$,
where $f_{\rm eq}(y')$ is the relativistic Fermi--Dirac distribution of some temperature which, for simplicity, we shall take to be the QED plasma temperature~$z$.%
\footnote{It may desirable to use the neutrino temperature estimated during run time instead of the QED plasma temperature. In practice, however, we find no significant difference in the outcome $\Neff$ between these choices.}
Then, the repopulation coefficient due to neutrino--neutrino collisions, $\{R_{\nu\nu}(y)\}_\alpha \equiv 2 \{\Gamma_y^{\nu\nu}\}_{\alpha \alpha}$,
can be constructed from the collision integral~\eqref{eq:collintnunu} to give
\begin{equation}
\begin{aligned}
\big\{R_{\nu \nu}(y) \big\}_{\alpha} = \, & \frac{2 G_F^2}{ (2 \pi)^3 y^2}
\int {\rm d}y_2 {\rm d}y_3 \,
\left[\Pi_{2a}^{\rm sc}(y, y_2) + 2 \Pi_{2b}^{\rm sc}(y, y_4) \right]\\
& \qquad \qquad \times \Big( [1-f_{\rm eq}(y_2)] f_{\rm eq}(y_3) f_{\rm eq}(y_4) +f_{\rm eq}(y_2) [1-f_{\rm eq}(y_3)][1-f_{\rm eq}(y_4)] \Big)\\
\equiv \, & {\cal D}(y,z),
\label{eq:selfdamping}
\end{aligned}
\end{equation}
which is {\it flavour-blind} as expected (because in thermal equilibrium, there are equally many neutrinos and antineutrinos in all flavours).

\begin{figure}[t]
\begin{center}
\includegraphics[width=10cm]{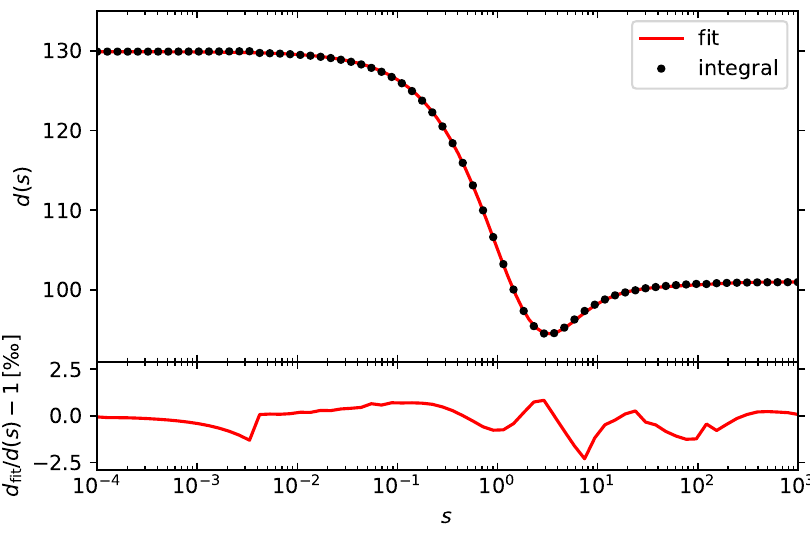}
\end{center}
\caption{The function $d(s)$, as defined in equation~\eqref{eq:dy}. The dots denote actual numerical evaluations of the integral~\eqref{eq:selfdamping}, while the line represents the fitting function $d_{\rm fit}(s)$ of equation~\eqref{eq:fittingfunction}. The bottom panel shows the factional difference between the two.\label{fig:dy}}
\end{figure}

For relativistic Fermi--Dirac distributions, the function ${\cal D}(y,z)$ evaluates to
\begin{equation}
{\cal D}(y,z) = \frac{2 G_F^2 y z^4}{(2 \pi)^3} d(y/z),
\label{eq:dy}
\end{equation}
which is predominantly linear in $y$, with a residual $y$-dependence encapsulated in the function $d(s=y/z)$, shown in figure~\ref{fig:dy} as a function of $s=y/z$. For computational ease, $d(s)$ can be fitted in the interval $s \in [10^{-4},10^3]$ to better than 0.25\% accuracy by the curve
\begin{equation}
d_{\rm fit}(s) = d_0 e^{-1.01 s} + d_\infty (1- e^{-0.01 s}) + (e^{-0.01 s}-e^{-1.01 s}) \left[\frac{a_0 + a_1 \ln (s) + a_2 \ln^2(s)}{1 + b_1 \ln (s) + b_2 \ln^2(s)} \right],
\label{eq:fittingfunction}
\end{equation}
where $d_0 = 129.875$ and $d_\infty = 100.999$ are the asymptotic values of the function as $s \to 0$ and
$s \to \infty$ respectively, and the fitting coefficients are $a_0 = 90.7332$, $a_1 = -48.4473$, $a_2 =20.1219$,
$b_1 = -0.529157$, and $b_2 = 0.20649$.

\paragraph{Neutrino--electron.} The repopulation coefficients corresponding to the neutrino--electron collision integral~${\cal I}_{\nu e}[\varrho(y)]$,
$\{R_{\nu e}(y)\}_\alpha \equiv 2 \{\Gamma_y^{\nu e}\}_{\alpha \alpha}$,
can be established similarly under the assumption of $\varrho(y') = f_{\rm eq}(y') \mathbb{1}$ for all $y'\neq y$,
where, again, $f_{\rm eq}(y')$ is the relativistic Fermi--Dirac distribution with the QED plasma temperature~$z$. We likewise assume the electron phase space distribution to be given by the same relativistic Fermi--Dirac form, i.e., $f_e(y) = f_{\rm eq}(y)$. Then, using equation~\eqref{eq:collint} to construct $\{\Gamma_y^{\nu e}\}_{\alpha \alpha}$, we find
\begin{equation}
\begin{aligned}
\{R_{\nu e}(y)\}_{\alpha} = \, & \frac{G_F^2}{ 2(2 \pi)^3 y^2} \, \left[(2 \sin^2 \theta_W \pm 1)^2_\alpha + 4 \sin^4 \theta_W\right] \int {\rm d}y_2 {\rm d}y_3 \, \left[\Pi_{2a}^{\rm sc}(y, y_2) + 2 \Pi_{2b}^{\rm sc}(y, y_4) \right]\\
& \qquad \qquad \times
\Big( [1-f_{\rm eq}(y_2)] f_{\rm eq}(y_3) f_{\rm eq}(y_4) +f_{\rm eq}(y_2) [1-f_{\rm eq}(y_3)][1-f_{\rm eq}(y_4)] \Big) \\
= \, & \frac{1}{4} \left[(2 \sin^2 \theta_W \pm 1)^2_\alpha + 4 \sin^4 \theta_W\right] {\cal D}(y,z),
\label{eq:nuedampingdiag}
\end{aligned}
\end{equation}
where in the prefactor $(2 \sin^2 \theta_W\pm 1)_\alpha$ the plus sign ``$+$'' is understood to apply to $\alpha =e$ and ``$-$'' to $\alpha = \mu, \tau$, and ${\cal D}(y,z)$ is the same function given in equation~\eqref{eq:dy}.

It is of course also possible to retain a finite $x$, i.e., a finite electron mass~$m_e$, in the computation of the repopulation and hence damping coefficients. However, the final outcome will have a more complicated time dependence than that contained in ${\cal D}(y,z)$, and for simplicity we have opted to omit this additional dependence.


\bibliography{bib}

\end{document}